\documentclass[aps,twocolumn,nofootinbib]{revtex4}

\usepackage{graphicx}% Include figure files
\usepackage{dcolumn}% Align table columns on decimal point
\usepackage{bm}% bold math
\usepackage{amssymb,amsfonts}
\usepackage{epsfig}
\usepackage{epstopdf}
\usepackage[all]{xy}
\usepackage{amsthm}
\usepackage{dcolumn}
\usepackage{hyperref}
\usepackage{url}

\newcommand{\be}{\begin{equation}}
\newcommand{\ee}{\end{equation}}
\newcommand{\bea}{\begin{eqnarray}}
\newcommand{\nn}{\nonumber}
\newcommand{\eea}{\end{eqnarray}}

\def\inbar{\,\vrule height1.5ex width.4pt depth0pt}
\def\IR{\relax{\rm I\kern-.18em R}}
\def\IC{\relax\hbox{$\inbar\kern-.3em{\rm C}$}}

\begin{document}

\title{Covariant and infrared-free graviton two-point function in de Sitter spacetime}

\author{Hamed Pejhan\footnote{h.pejhan@piau.ac.ir}}
\affiliation{Department of Physics, Science and Research Branch, Islamic Azad University, Tehran, Iran}

\author{Surena Rahbardehghan\footnote{s.rahbardehghan@iauctb.ac.ir}}
\affiliation{Department of Physics, Islamic Azad University, Central Tehran Branch, Tehran, Iran}

\begin{abstract}
In this paper, the two-point function of linearized gravitons on de Sitter (dS) space is presented. Technically, respecting the dS ambient space notation, the field equation is given by the coordinate-independent Casimir operators of the de Sitter group. Analogous to the quantization of the electromagnetic field in Minkowski space, the field equation admits gauge solutions. The notation allows us to exhibit the formalism of Gupta-Bleuler triplets for the present field in exactly the same manner as it occurs for the electromagnetic field. In this regard, centering on the spin-two part (the traceless part, ${\cal{K}}^t$), the field solution is written as a product of a generalized polarization tensor and a minimally coupled massless scalar field. Then, admitting a de Sitter-invariant vacuum through the so-called "Krein space quantization", the de Sitter fully covariant two-point function is calculated. This function is interestingly free of pathological large distance behavior (infrared divergence). Moreover, the spin-zero part (the pure-trace part; ${\cal{K}}^{pt}$) of the field is discussed in this paper. It is shown that the implications of the dS group unitary irreducible representations restrict the gauge-fixing parameter to the optimal value, which remarkably results in the pure-trace part be written in terms of a conformally coupled massless scalar field.
\end{abstract}
\maketitle
%\tableofcontents

\section{Introduction}
\label{sec:intro}
The recent cosmological observations are strongly in favor of a positive acceleration of the present Universe \cite{riess}, so that, with good accuracy the background spacetime can be considered a de Sitter space. Therefore, the study of the linear perturbations of Einstein gravity around the de Sitter metric (the dS linear quantum gravity) and the associated two-point function are of particular interest. However, the infrared (IR) properties of the dS graviton two-point function have remained a source of controversy over the past 30 years. The graviton two-point function in de Sitter space behaves in a manner similar to that for the minimally coupled massless scalar field \cite{Ford1601}, for which there is no Hilbertian dS-invariant vacuum state because of infrared divergences \cite{Allen3136,Allen3771}. This similarity leads to infrared divergences in the graviton two-point function \cite{Allen813,Floratos373,Antoniadis1037}. However, contrary to infrared divergences relevant to those of the massless scalar field theory in Minkowski space, it has been shown that there are no physical IR divergences in the graviton two-point function in de Sitter spacetime \cite{Ford1601}. Indeed, it is proved that the IR divergence of the graviton propagator on a de Sitter background does not manifest itself in the quadratic part of the effective action in the one-loop approximation \cite{Antoniadis437}.

From another perspective, the IR divergences have been considered by some authors to create instability in the de Sitter space \cite{Ford,Antoniadis1319}. Accordingly, in terms of the dS flat coordinate, the field operator for linear gravity has been investigated by Tsamis and Woodard, and they have examined the resulting possibility of quantum instability \cite{Tsamis217}. Such a quantum field, however, breaks the de Sitter invariance (the coordinate covers only one-half of the de Sitter hyperboloid).

Recently, however, by admitting a de Sitter-invariant vacuum in an indefinite inner product space, it has been proved that a causal and fully covariant construction of the minimally coupled massless\footnote{It is worth mentioning that, in a generic curved spacetime, there is no obvious definition of the mass concept. However, in the (anti-)de Sitter spacetime thanks to the maximal symmetry of these spaces, the mass concept can be defined precisely \cite{Gazeau304008,Flato415}.} scalar quantum field on de Sitter spacetime can be structured into the so-called "Krein space quantization" \cite{de Bievre6230,Gazeau1415}. The causality and the covariance of the theory are actually assured thanks to a suitable choice of the space of solutions of the classical field equation. Interestingly, contrary to what happened in previous treatments of this problem, the formalism suffers from neither infrared nor ultraviolet divergences (this formalism yields an automatic covariant renormalization of the energy-momentum tensor). The Krein space quantization, therefore, provides a proposal to construct the covariant and infrared-free graviton two-point function on a de Sitter background \cite{Behroozi124014,Dehghani064028,Garidi032501,Pejhan2052,Rahbardehghan}.

In this paper, respecting these capabilities and also considering a rigorous group theoretical approach, the dS linear quantum gravity (neglecting the graviton-graviton interactions) is investigated. At the beginning, in Sec. II, the de Sitter graviton field equation is presented. Then, the dS ambient formalism (five global coordinates) and the two independent Casimir operators of the dS group are introduced. Utilizing the ambient space notation provides the opportunity to express the field equation in terms of the coordinate-independent de Sitter Casimir operators. These operators enable one to classify the unitary irreducible representations (UIRs) of the dS group \cite{Dixmier9,Takahashi289}. On this basis, it is discussed that the field equation admits gauge solutions, and one is free to consider a gauge-fixing parameter $c$.

Therefore, in Sec. III, with regard to the field equation, constrained with the tracelessness condition (the pure-trace part will be discussed in Sec. VII), we define a Gupta-Bleuler triplet to manage the covariance and the gauge invariance of the theory. Indeed, as discussed in Ref. \cite{BinegarGUPTA}, "the appearance of the Gupta-Bleuler triplet seems to be universal in gauge theories, and crucial for quantization". Thanks to the ambient space notation, an exhibition of the Gupta-Bleuler triplet for our considered field occurs in exactly the same manner as the electromagnetic field (for more mathematical detail, see for instance \cite{Garidi032501}). Accordingly, it is pointed out that the invariant space is determined by an indecomposable representation of the dS group, while, physical states (the central part of the indecomposable representation) correspond to the UIR's ${\Pi}^{\pm}_{2,2}$ (the Dixmier's notation \cite{Dixmier9}).

Section IV is devoted to the solution of the traceless part of the field equation, which is a coordinate-independent function (thanks to the dS ambient notation). We express the field solution ${\cal{K}}^t$ in terms of a polarization tensor and the minimally coupled massless scalar field:
$${\cal{K}}^t_{\alpha\beta}(x)= {\cal{D}}_{\alpha\beta}(x,\partial)\phi(x).$$
Actually, it is shown that quantizing entails preliminary covariant quantization of the minimally coupled massless scalar field. For which, as mentioned, it has been claimed that there is no Hilbertian de Sitter-invariant vacuum state, so no covariant Hilbert space quantization is possible \cite{Allen3136,Allen3771}.

Interestingly, a closer look at the situation reveals a further Gupta-Bleuler triplet (the Krein-Gupta-Bleuler triplet) lying behind the minimally coupled massless scalar field \cite{de Bievre6230,Gazeau1415}. More precisely, it is proved that a rather straightforward application of the Gupta-Bleuler formalism, without changing the physical content of the theory, permits one to avoid the symmetry-breaking altogether; the constructed field transforms correctly under de Sitter and gauge transformations and acts on a state space containing a vacuum invariant under all of them. It is free of infrared divergence \cite{de Bievre6230,Gazeau1415}. A brief discussion of these statements is given in Sec. V. Consistency of the Krein quantization method with the usual ones in the Minkowskian limit, as a necessity of any successful quantization scheme in dS space, and the unitarity condition are also studied in this section.

In Sec. VI, the Krein space quantization method is utilized to calculate the dS covariant two-point function for the spin-two sector, ${\cal{W}}^t_{\alpha\beta\alpha'\beta'}(x,x')$, in the ambient space notation. It is also written in terms of the intrinsic coordinate. This function fulfils the conditions of (1) locality, (2) covariance, (3) transversality, (4) permutational index symmetries and (5) tracelessness. Interestingly, it is free of any infrared divergences.

In Sec. VII the pure-trace part (the so-called conformal sector) of the tensor field ${\cal{K}}^{pt}$ is studied. It is proved that the gauge-fixing procedure, based on the group theoretical impositions, results in the pure-trace part be written in the sense of a conformally coupled massless scalar field.

Finally, a brief summery and discussion are given in Sec. VIII. Some mathematical details of calculations are supplied in the Appendixes.

\section{De Sitter linear quantum gravity: the field equation}
\label{sec:equiv}
In this section, by splitting our metric into a dS fixed background $g_{\mu\nu}^{dS}$ and a small fluctuation $h_{\mu\nu}$, i.e. $g_{\mu\nu}=g_{\mu\nu}^{dS}+h_{\mu\nu}$, we briefly study the linearized graviton field equation in de Sitter spacetime. At this linear approximation, the reparametrization invariance implies the following gauge invariance:
\begin{equation}\label{2.4} h_{\mu\nu}\rightarrow h_{\mu\nu} + 2\nabla_{(\mu}\Xi_{\nu)},\end{equation}
where $\Xi_{\nu}$ is an arbitrary vector field and $\nabla_{(\mu}\Xi_{\nu)}= \frac{1}{2} (\nabla_{\mu}\Xi_{\nu} + \nabla_{\nu}\Xi_{\mu})$. No term proportional to $h^2$ appears in the linearized Lagrangian, therefore, the propagating dS wave equation for tensor fields $h_{\mu\nu}$ is \cite{Higuchi3077,Fronsdal848}
\begin{eqnarray} \label{2.3}
(\Box_H + 2H^2)h_{\mu\nu} - (\Box_H - H^2)g_{\mu\nu}^{dS}h' - 2\nabla_{(\mu}\nabla^\rho h_{\nu)\rho} \nonumber \\ + g_{\mu\nu}^{dS}\nabla^\lambda\nabla^\rho h_{\lambda\rho} + \nabla_\mu\nabla_\nu h'=0,
\end{eqnarray}
$H$ refers to the Hubble constant, $\nabla^\nu$ to the dS covariant derivative, $\Box_H= g_{\mu\nu}^{dS}\nabla^\mu\nabla^\nu$ to the Laplace-Beltrami operator and $h'=(g^{dS})^{\mu\nu}h_{\mu\nu}$. Here, we consider a generalization of the Lorentz gauge condition, i.e.,
\begin{equation}\label{2.5}
\nabla^\mu h_{\mu\nu} = \zeta\nabla_\nu h',
\end{equation}
where $\zeta$ is an arbitrary constant. Technically, by adding a gauge-fixing term to the Lagrangian, one can fix the gauge. In this regard, the ambient notation is introduced in the next part to express the field equation (\ref{2.3}) in terms of the coordinate-independent de Sitter-Casimir operators. Interestingly, it will be shown that considering the specific value $\zeta=\frac{1}{2}$, the relation between the tensor field and the dS group representations becomes apparent.

\subsection{The de Sitter ambient space notation}
De Sitter spacetime can be considered as a one-sheeted hyperboloid embedded in a 5-dimensional Minkowski space ($\alpha,\beta= 0,1,2,3,4$)
\begin{eqnarray}\label{2.1}
{X_H}  = \{ x \in {R}^5 ; x^2={\eta}_{\alpha\beta} {x^{\alpha}} {x^\beta} =  -H^{-2}\},
\end{eqnarray}
where $\eta_{\alpha\beta}=$ diag$(1,-1,-1,-1,-1)$. The dS metric is ($\mu,\nu=0,1,2,3$)
\begin{equation}\label{2.2}
ds^2=\eta_{\alpha\beta}dx^{\alpha}dx^{\beta}|_{x^2=-H^{-2}}=g_{\mu\nu}^{dS}dX^{\mu}dX^{\nu}.
\end{equation}
$x^\alpha$ and $X^\mu$ are, respectively, referred to the dS ambient space formalism (five global coordinates) and the dS intrinsic coordinate (four local coordinates).

Here, we consider the ambient space coordinate, which enables us to formulate the dS elementary systems (in the Wigner sense) in analogy with the Minkowskian case. In this notation, the tensor field ${\cal{K}}_{\alpha\beta}(x)$ must satisfy the following conditions:
\begin{itemize}
\item{homogeneity \begin{equation}\label{2.6} x^\alpha \frac{\partial}{\partial x^\alpha}{\cal{K}}_{\beta\gamma}(x) = x\cdot\partial {\cal{K}}_{\beta\gamma}(x) = \sigma{\cal{K}}_{\beta\gamma}(x).\end{equation}}

\item{transversality \begin{equation}\label{2.7} x^\alpha {\cal{K}}_{\alpha\beta}(x) = x^\beta {\cal{K}}_{\alpha\beta}(x)\;\Big(\equiv x\cdot {\cal{K}}(x)\Big) =0. \end{equation}}
\end{itemize}
In this coordinate, the covariant derivative is
\begin{equation}\label{2.8}
D_\beta T_{\alpha_1...\alpha_i...\alpha_n}=\bar{\partial}_\beta T_{\alpha_1...\alpha_i...\alpha_n}-H^2\sum_{i=1}^n x_{\alpha_i}T_{\alpha_1...\beta...\alpha_n},
\end{equation}
where $\bar{\partial}$ is the tangential (or transverse) derivative
\begin{equation}\label{2.9}
\bar{\partial}_\alpha=\theta_{\alpha \beta}\partial^\beta=\partial_\alpha+H^2x_\alpha x\cdot\partial,\;\;\;\; x\cdot\bar{\partial}=0.
\end{equation}
$\theta_{\alpha \beta}=\eta_{\alpha \beta}+H^2x_\alpha x_\beta$ is the transverse projector. It is the only symmetric and transverse tensor which is related to the de Sitter metric
$$g_{\mu\nu}^{dS}=\frac{\partial x^\alpha}{\partial X^\mu}\frac{\partial x^\beta}{\partial X^\nu}\theta_{\alpha \beta}.$$
The "intrinsic" field $h_{\mu\nu}(X)$ is locally determined by the "transverse" tensor field ${\cal{K}}_{\alpha\beta}(x)$ through
\begin{equation}\label{2.10}
h_{\mu\nu}(X)=\frac{\partial x^\alpha}{\partial X^\mu}\frac{\partial x^\beta}{\partial X^\nu}{\cal{K}}_{\alpha\beta}\big( x(X) \big).
\end{equation}
Accordingly, to investigate the relation between the tensor field and the irreducible representations of the dS group, one can easily write the field equation (\ref{2.3}) in terms of the dS Casimir operator.

The de Sitter kinematical group is the $10$-parameter group $SO_0(1,4)$ (connected component of the identity in $O(1,4)$). It has two Casimir operators\footnote{The subscript 2 in $Q_2^{(1)}$ and $Q_2^{(2)}$ reminds that the carrier space is constituted by second rank tensors.}
\begin{equation} \label{2casimir} Q_2^{(1)}=-\frac{1}{2}L^{\alpha\beta}L_{\alpha\beta},\;\;\;\;\ Q_2^{(2)}=-W_{\alpha}W^{\alpha},\end{equation}
$W_{\alpha}=-\frac{1}{8}\epsilon_{\alpha\beta\gamma\sigma\eta}L^{\beta\gamma}L^{\sigma\eta}$ and $\epsilon_{\alpha\beta\gamma\sigma\eta}$ is the antisymmetric tensor in the ambient space notation with $\epsilon_{01234}=1$. The dS group generator is $L_{\alpha\beta}=M_{\alpha\beta}+{{\sum}_{\alpha\beta}}$, in which \cite{Gazeau2533}
\begin{eqnarray}
M_{\alpha\beta}\equiv-i(x_{\alpha}\partial_{\beta}-x_{\beta}\partial_{\alpha})=-i(x_{\alpha}\bar\partial_{\beta}-x_{\beta}\bar\partial_{\alpha}),
\end{eqnarray}
and
\begin{eqnarray}
{\sum}_{\alpha\beta}{\cal{K}}_{\gamma\delta ...}\equiv-i(\eta_{\alpha\gamma}{\cal{K}}_{\beta\delta ...}\hspace{3.5cm}\nn\\
-\eta_{\beta\gamma} {\cal{K}}_{\alpha\delta ...} + \eta_{\alpha\delta}{\cal{K}}_{\gamma\beta ...}-\eta_{\beta\delta} {\cal{K}}_{\gamma\alpha ...} + ...).
\end{eqnarray}
On this basis, the action of $Q_2(\equiv  Q_2^{(1)})$ on ${\cal{K}}$ can be written explicitly as
\begin{equation}\label{2.19} Q_2{\cal{K}}=(Q_0-6){\cal{K}}+2\eta{\cal{K}}'+2{\cal{S}}x\partial\cdot{\cal{K}}-2{\cal{S}}\partial x\cdot{\cal{K}}, \end{equation}
$Q_0=-\frac{1}{2}M_{\alpha\beta}M^{\alpha\beta}=-H^{-2}(\bar{\partial})^2$ is the scalar Casimir operator and ${\cal{S}}$ is the symmetrizer operator (${\cal{S}}\xi_\alpha \omega_\beta=\xi_\alpha \omega_\beta+\xi_\beta \omega_\alpha$).

Accordingly, the field equation for ${\cal{K}}$  takes the following form \cite{Dehghani064028,Fronsdal848},
\begin{equation}\label{2.21} (Q_2+6){\cal{K}}(x)+D_2\partial_2\cdot{\cal{K}}(x)=0, \end{equation}
in which the operator $D_2$ is defined by
\begin{equation}\label{2.15} D_2K=H^{-2}{\cal{S}}(\bar{\partial}-H^2x)K, \end{equation}
and $\partial_2\cdot$, the generalized divergence on the de Sitter hyperboloid, is as follows,
\begin{equation}\label{2.16} \partial_2\cdot{\cal{K}}=\partial\cdot{\cal{K}}-H^2x{\cal{K}}'-\frac{1}{2}H^2D_1{\cal{K}}', \end{equation}
$D_1=H^{-2}\bar{\partial}$ and ${\cal{K}}'$ is the trace of ${\cal{K}}_{\alpha\beta}$. Considering an arbitrary vector field ${\Lambda_g}$ and the following relations \cite{Garidi3838,Gazeau5920},
\begin{eqnarray}\label{identities}
\partial_2\cdot D_2{\Lambda_g}=-(Q_1+6){\Lambda_g},\;\; Q_2D_2{\Lambda_g}=D_2Q_1{\Lambda_g},\nn\\
Q_1{\Lambda_g}=(Q_0-2){\Lambda_g}+2x \partial\cdot{\Lambda_g}-2\partial x\cdot{\Lambda_g},\hspace{0.5cm}
\end{eqnarray}
one can simply show that Eq. (\ref{2.21}) is invariant under the following gauge transformation \cite{Gazeau+yo},
\begin{equation}\label{gauge} {\cal{K}}\rightarrow{\cal{K}}+D_2{\Lambda_g}. \end{equation}
In this notation, the general gauge condition (\ref{2.5}) would be
\begin{equation}\label{2.23}
\partial_2\cdot{\cal{K}}=(\zeta-\frac{1}{2})\bar{\partial}{\cal{K}}'.
\end{equation}

Consistently with (\ref{2.21}), the following action can be considered
\begin{equation}\label{2.22}
S=\int d\sigma{\cal{L}},\;\;{\cal{L}}=-\frac{1}{2x^2}{\cal{K}}\cdot\cdot(Q_2+6){\cal{K}}+\frac{1}{2}(\partial_2\cdot{\cal{K}})^2.
\end{equation}
$d\sigma$ is the volume element in dS space. By adding a gauge-fixing term to the Lagrangian, one obtains \cite{Gazeau+yo}
\begin{eqnarray}\label{2.24}
{\cal{L}} & = & -\frac{1}{2x^2}{\cal{K}}\cdot\cdot(Q_2+6){\cal{K}}+\frac{1}{2}(\partial_2\cdot{\cal{K}})^2 \nn\\ &+&\frac{1}{2\alpha}\Big( \partial_2\cdot{\cal{K}}-(\zeta-\frac{1}{2})\bar{\partial}{\cal{K}}' \Big)^2.
\end{eqnarray}
Then, by choosing $\zeta=\frac{1}{2}$, we have ($c=\frac{1+\alpha}{\alpha}$)
\begin{equation}\label{2.25} {\cal{L}}=-\frac{1}{2x^2}{\cal{K}}\cdot\cdot(Q_2+6){\cal{K}}+\frac{c}{2}(\partial_2\cdot{\cal{K}})^2.\end{equation}
\begin{equation}\label{2.26} (Q_2+6){\cal{K}}(x)+cD_2\partial_2\cdot{\cal{K}}(x)=0.\end{equation}
In analogy with the electromagnetic field in Minkowski space, the field equation admits gauge solutions ($c$ is the gauge-fixing parameter). Quantizing gauge-invariant theories, as is well known, usually requires a quantization scheme \`{a} \emph{la} Gupta-Bleuler \cite{BinegarGUPTA,GazeauGUPTA}. It has, in fact, been proved that the use of an indefinite metric is an unavoidable feature if one insists on the preserving of causality (locality) and covariance in gauge quantum field theories.

Before coming back to this point, the group-theoretical content of the field equation will be described in the following part.

\subsection{De Sitter group interpretation}
Generally, the tensor field is composed of two parts:
\begin{equation}\label{p1} {\cal{K}}_{\alpha\beta}(x) = {\cal{K}}_{\alpha\beta}^t(x)+{\cal{K}}_{\alpha\beta}^{pt}(x). \end{equation}
The spin-two part (the traceless part) of the theory $ {\cal{K}}_{\alpha\beta}^t$ and the spin-zero part (the pure-trace part) ${\cal{K}}_{\alpha\beta}^{pt}$. It is worth mentioning that the pure-trace part does not correspond to a UIR of the dS group. Indeed, the traceless conditions on the tensor field is a necessary condition in order to relate it to the UIRs of the dS group \cite{Gazeau5920}. In the other word, in the context of general relativity, the pure-trace part of the graviton field does not carry any dynamics. However, when matter quantum fields are taken into account, this part of the metric attains a dynamical content. Actually the pure-trace part is considerable for establishing inflationary scenarios of the universe. A scalar field called inflaton is introduced in these models because of that, the pure-trace part of the metric becomes dynamical and it must be quantized \cite{Antoniadis2013,Antoniadis4770}. Through this process, then, a gravitational instability is produced that combined with the primordial quantum fluctuations of the inflaton scalar field define the inflationary model. These models are capable of describing the configuration of the galaxies, clusters of galaxies and the large scale structure of the universe \cite{Lesgourgues}. The pure-trace part will be considered in Sec. VII.

The spin-two part, the traceless massless tensor field ${\cal{K}}_{\alpha\beta}^{t}$, however, corresponds to the associated indecomposable representation of the de Sitter group. Actually, with regard to the definition of the dS group Casimir operators for the spin-2 tensor representations relevant to the present work, the operator $Q_2(\equiv  Q_2^{(1)})$ commutes with the action of the group generators and, as a consequence, it is constant in the corresponding UIR. Hence, the UIR's can be classified by considering the eigenvalues of $Q_2$, i.e., $\langle Q_2\rangle$, as
\begin{equation}\label{2.27} (Q_2 - \langle Q_2\rangle){\cal{K}}^t(x)=0. \end{equation}
According to the Takahashi and Dixmier's notation \cite{Dixmier9,Takahashi289}, the eigenvalues of the Casimir operator are classified under the following series representations (for a detailed discussion about the relevant representations, one can refer to \cite{Garidi3838}):
\begin{itemize}
\item{Principal series representations $(U^{2,\nu})$, also called "massive" representations \cite{Gazeau304008,Flato415},
\begin{equation}\label{2.30} \langle Q_2\rangle = \nu^2 - \frac{15}{4},\;\; \nu\in\mathbb{R}.  \end{equation}
This series of UIR's admits a massive Poincar\'{e} group UIR in the limit $H=0$.}
\item{Complementary series representations $(V^{2,\mu})$,
\begin{equation}\label{2.31} \langle Q_2\rangle = \mu - 4, \;\; \mu\in\mathbb{R},\;0<\mu<\frac{1}{4}.  \end{equation}}
\item{Discrete series representations $({\Pi}^{\pm}_{2,q})$, also called the "massless" representations \cite{Gazeau304008,Flato415},
\begin{equation}\label{2.32} \langle Q_2\rangle = - 6 - (q+1)(q-2), \;\; q={1},{2}.  \end{equation}}
\end{itemize}
For the discrete series, considering the parameter $q=1$ ($\langle Q_2\rangle =-4$), results in the representation ${\Pi}^{\pm}_{2,1}$, which does not have a corresponding counterpart in the Minkowskian limit. The second value, $q=2$ ($\langle Q_2\rangle =-6$), results in the representation ${\Pi}^{\pm}_{2,2}$. They are precisely the unique extensions of the massless Poincar\'{e} group representations with helicity $\pm2$.

Now, comparing the field equation (\ref{2.26}) with (\ref{2.27}) reveals that the traceless part of the solution of (\ref{2.26}) transforms under indecomposable representations (as opposed to irreducible representations) of the dS group. More precisely, the physical states are related to a subspace specified by the divergencelessness condition imposed on the field operator, while the field operator must be built on a larger gauge dependent space which is defined with the associated indecomposable representation of the de Sitter group. The physical states correspond to the UIR's ${\Pi}^{\pm}_{2,2}$. It is the central part of the indecomposable representation (for a detailed discussion, see the next section).

Indeed, respecting the dS physical representations, massive elementary systems are associated with unitary irreducible representations of the dS kinematic group \cite{Garidi3838}. While, massless elementary systems are connected to the indecomposable representations of this group \cite{Garidi032501,Gazeau329}. As usual they display gauge invariance and conformal invariance properties. Thereby, quantizing the massless tensor fields necessitates the fixing of the gauge. On curved backgrounds, however, finding the optimal value of $c$ is a nontrivial question. Respecting the physical representations of the de Sitter group, it has been claimed that the specific choice $c = 2/(2s + 1)$ ($s$ is the angular momentum, spin, of the field) restricts the space of solutions to the minimal content of any massless invariant theory \cite{Gazeau2533,Gazeau507}. More precisely, any other choice of $c$ introduces logarithmic singularities, which implies reverberation inside the light cone \cite{Gazeau329}. In this regard, the massless vector field in a de Sitter universe has been investigate in Ref. \cite{Garidi032501}. It is shown that $c = 0$, in contrast to the Minkowskian limit (Feynman gauge), is not the minimal (or optimal) choice. It actually yields logarithmic divergent terms in the vector field expression. Consistently with the general formula $c = 2/(2s + 1)$, it is proved that the minimal choice is $c=\frac{2}{3}$.

According to the above statements, in the case of the massless spin-2 fields, it is expected that the optimal choice, for which the logarithmic contribution disappears, is $c=\frac{2}{5}$. Nevertheless, in Sec. IV we explicitly prove that by applying an extra condition, $\bar\partial\cdot K= 0$ (see Eq. (\ref{4.3}) and its following identities), no logarithmic divergence appears in the field equation. Therefore, the gauge fixing parameter does not need to be fixed to $2/5$. [In order to derive the identities listed in Appendix B, this condition is required. Actually Eq. (\ref{4.14}) is calculated through this very condition.]

Here, we should emphasize that our result is not in contradiction with the representations of the dS group. In our work, applying the aforementioned condition contracts the space of solutions without losing the minimal requirements for the solutions. The gauge fixing parameter, however, must be set to $2/5$ when the solution space is intact to remove the logarithmic singular terms. By relaxing this condition, the solving procedure is more complicated. It is expected that it yields $c=\frac{2}{5}$ \cite{Gazeau329}.

Pursuing our calculations, through relaxing the gauge-fixing parameter $c$ (it does not need to be fixed to $2/5$), provides a nontrivial remarkable advantage for the theory, especially, when the pure-trace part (conformal sector) of the field is taken into account. It will be discussed in Sec. VII.

\section{The Gupta-Bleuler triplet}
In this section, considering the graviton field equation (\ref{2.26}) constrained with the tracelessness condition ${\cal{K}}'=0$, the Gupta-Bleuler triplet $V_g\subset V\subset V_c$ carrying the indecomposable structure of the related dS UIRs is introduced. In this regard, we consider $V_c$ as the space of all square integrable solutions of the field equation, respecting the following dS-invariant (indefinite) inner product \cite{Gazeau329},
\begin{eqnarray} \label{3.1}
({\cal{K}}^t_1,{\cal{K}}^t_2)&=& \frac{i}{H^2}\int_{S^3,\rho=0} [({\cal{K}}_1^t)^*\cdot\cdot\partial_\rho{\cal{K}}^t_2 \nn\\
&-& 2c(\partial_\rho x\cdot{({\cal{K}}_1^t)}^*\cdot\cdot(\partial\cdot{\cal{K}}^t_2) - (1^* \leftrightharpoons 2)]d\Omega,\hspace{0.9cm}
\end{eqnarray}
where ${\cal{K}}^t_1$ and ${\cal{K}}^t_2$ are two different modes on the solutions space, and "$\cdot\cdot$" is a shortened notation for total contraction. Here, the system of bounded global intrinsic coordinates $(X^\mu,\;\mu=0,1,2,3)$, well suited to characterize a de Sitter compactified version (i.e. $S^3\times S^1$), is used
\begin{eqnarray}\label{3.2} \left \{\begin{array}{rl}
&x^0 = H^{-1}\tan \rho, \vspace{2mm}\\
&x^1 = (H\cos\rho)^{-1}(\sin\alpha\sin\theta\cos\varphi),\vspace{2mm} \\
&x^2 = (H\cos\rho)^{-1}(\sin\alpha\sin\theta\sin\varphi),\vspace{2mm}\\
&x^3 = (H\cos\rho)^{-1}(\sin\alpha\cos\theta),\vspace{2mm}\\
&x^4 = (H\cos\rho)^{-1}(\cos\alpha),
\end{array}\right.
\end{eqnarray}
where $-\pi/2<\rho<\pi/2,\; 0\leq\alpha\leq\pi, \; 0\leq\theta\leq\pi$ and $0\leq\varphi<2\pi$.

The physical states verify the divergencelessness condition and belong to an invariant subspace of the solutions, the space $V$, for which the inner product is \cite{Gazeau329}
\begin{eqnarray}
({\cal{K}}^t_1,{\cal{K}}^t_2)= \frac{i}{H^2}\int_{S^3,\rho=0} [({\cal{K}}^t_1)^*\cdot\cdot\partial_\rho{\cal{K}}^t_2\hspace{3cm}\nn\\
- {\cal{K}}^t_2\cdot\cdot\partial_\rho({\cal{K}}_1^t)^*]d\Omega.\hspace{1cm}
\end{eqnarray}
Contrary to $V_c$, it is obviously $c$ (gauge) independent. The space of gauge solutions, $K_g = D_2{\Lambda_g}$, is denoted by an invariant subspace $V_g$ of $V$. These are orthogonal to every element in $V$ including themselves. The inner product is semidefinite in $V$ and is positive definite in the quotient space $V/V_g$. The dS group acts on $V/V_g$ through the massless, helicity $\pm 2$ unitary representation $\Pi^+_{2,2}\oplus\Pi^-_{2,2}$. It is indeed the physical states space. Here, we must underline that all three of these spaces carry representations of the dS group, but $V$ and $V_g$ are not invariantly complemented.

Now, the gauge state space $V_g$, the vector states $\partial_2\cdot{\cal{K}}^t$ belonging to $V_c/V$, and the physical states space $V/V_g$ should be characterized.

\subsection{The gauge states space}
Considering $K_g = D_2\Lambda_g$ and Eqs. (\ref{identities}), the field equation (\ref{2.26}) reduces to (constrained with the tracelessness condition ${\cal{K}}'=0$)
\begin{equation}\label{3.3} (1-c)D_2(Q_1+6)\Lambda_g=0.\end{equation}
Therefore, we have
\begin{itemize}
\item{For $c=1$, the vector field $\Lambda_g$ is unrestricted and bears merely the differentiability conditions. Then the gauge states is determined by $D_2\Lambda_g$.}
\item{For $c\neq 1$, it is clear that the space of solutions of (\ref{3.3}), possessing the divergencelessness ($\partial\cdot\Lambda_g=0$) and transversality ($x\cdot \Lambda_g=0$) conditions, carries a vector representation \cite{Gazeau5920,Garidi032501}. This vector field can be written as \cite{Gazeau5920}}
    \begin{equation}\label{3.4} \Lambda_g=\bar Z \phi_1 + D_1 \phi_2.\end{equation}
    where $Z$ is a constant five-vector field, and
    $$(Q_0 + 4)\phi_1=0,$$
    $$\phi_2= -\frac{1}{6}(2H^2 x\cdot Z\phi_1 + Z\cdot\bar\partial\phi_1).$$
    $\phi_1$ is demonstrated by the scalar representation of the dS group \cite{Gazeau1415}.
\end{itemize}

\subsection{The vector states space}
The vector states $\partial_2\cdot{\cal{K}}^t$ are characterized by
\begin{equation}\label{3.5'} \partial_2\cdot \Big( (Q_2+6){\cal{K}}^t(x)+cD_2\partial_2\cdot{\cal{K}}^t\Big)=0,\end{equation}
respecting the identities given in Eqs. (\ref{identities}) and $\partial_2{\cdot{Q_2} {\cal{K}}}^t = Q_{1} \partial\cdot{\cal{K}}^t$, so one can easily obtain
\begin{equation}\label{3.5} (1-c)(Q_1+6){\partial}_2\cdot {\cal{K}}^t=0.\end{equation}
Thus, we have
\begin{itemize}
\item{For $c=1$, it has no restriction with the exception of differentiability conditions.}
\item{For $c\neq 1$, it corresponds to a vector field similar to the gauge states.}
\end{itemize}

\subsection{The physical states space}
The physical states space, which is $c$-independent, is given by imposing the divergenceless condition on Eq. (\ref{2.26}) as follows:
\begin{equation}\label{3.6} (Q_2+6){\cal{K}}^t=0.\end{equation}

In the next section, the general solution of the field equation (\ref{2.26}) will be calculated in terms of a projection tensor field and the minimally coupled massless scalar field.

\section{The field solution (the spin-two sector)}
\label{sec:chapIV}
In the dS ambient formalism, the most general transverse, symmetric field ${\cal{K}}_{\alpha\beta}$ can be written in the following form \cite{Gazeau2533}
\be \label{4.1}
{\cal{K}}=\theta\phi_1+ {\cal{S}}\bar Z_{1}K+D_{2}K_{g},
\ee
in which $\phi_1$ is a scalar field, $K$ and $K_g$ are two transverse vector fields ($x\cdot K= 0 = x\cdot K_g$), $Z_1 (= Z_{1\alpha})$ is a five-dimensional constant vector ($\bar Z_{1\alpha}= \theta_{\alpha\beta}Z_1^\beta$). Imposing the traceless condition on (\ref{4.1}) yields
\begin{eqnarray}\label{4.2}
{\cal{K}}'  =  2\phi_1 + Z_{1}\cdot K + H^{-2}\bar\partial\cdot K_g = 0.
\end{eqnarray}

Then, by substituting $\cal{K}_{\alpha\beta}$ in (\ref{2.26}), we have
\begin{eqnarray}\label{4.3}
\left \{ \begin{array}{rl} (Q_0+6)\phi_1&=-4Z_1.K,\,\hspace{1.6cm}{(I)}\vspace{2mm}\\
\vspace{2mm} (Q_1+2)K &+\;cD_1 \partial\cdot K =0, \hspace{0.8cm} {(II)}\\
\vspace{2mm}(Q_1+6)K_g&= \frac{c}{2(c-1)} H^2 D_1 \phi_1 \\ \vspace{2mm}&+ \frac{2-5c}{1-c}H^2x\cdot Z_1 K \\ \vspace{2mm}&+ \frac{c}{1-c}(H^2xZ_1\cdot K \\ \vspace{2mm}& \hspace{1.2cm}- Z_1\cdot\bar\partial K),\hspace{0.4cm}(III)\end{array}\right.
\end{eqnarray}
Imposing an extra condition $\bar\partial\cdot K= 0$,\footnote{Note that, for transverse tensors like $K$; $\partial\cdot K = \bar\partial\cdot K$.} Eq. (\ref{4.3}-II) reduces to $(Q_1+2)K= Q_0K=0$, and using Eq. (\ref{4.3}-I), we obtain
\begin{equation} \label{4.4} \phi_1=-\frac{2}{3}Z_1.K, \;\;\;\;\;\;Q_0 \phi_1=0. \end{equation}
Considering Eqs. (\ref{4.2}) and (\ref{4.4}) then leads to
\begin{equation}\label{Kg} \bar\partial\cdot K_g =\frac{1}{3}H^2 Z_{1}\cdot K. \end{equation}

The general solution of (\ref{4.3}-II) would be \cite{Gazeau2533}
\begin{equation}\label{4.5} K=\bar Z_2 \phi_2+D_1 \phi_3,\end{equation}
in which $\phi_2$ and $\phi_3$ are two scalar fields, and $Z_2$ is another 5-dimensional constant vector. Substituting $K$ into (\ref{4.3}-II) results in
\begin{equation} \label{4.7} Q_0 \phi_2=0. \end{equation}
Therefore, the scalar field $\phi_2$ is a "massless" minimally coupled scalar field. With regard to the divergenceless condition ($\bar\partial\cdot K= 0$), one can then obtain
\begin{equation} \label{4.6} \phi_3 =-\frac{1}{2}[Z_2.\bar\partial\phi_2 + 2H^2x.Z_2\phi_2]. \end{equation}
So, the vector field $K$ can be written as
\begin{equation} \label{4.8} K=\bar Z_{2}\phi_2 - \frac{1}{2} D_1 [Z_2\cdot\bar\partial\phi_2 + 2H^2 x\cdot Z_2\phi_2],\end{equation}
and consequently
\begin{equation} \label{4.9} \phi_1 =-\frac{2}{3}{Z_1}\cdot \Big( \bar Z_{2}\phi_2 - \frac{1}{2}D_1[Z_2\cdot\bar\partial\phi_2 + 2H^2x\cdot Z_2\phi_2]\Big). \end{equation}

The vector field $K_g$, respecting (\ref{4.3}-III), after simple calculations (see Appendix B) can be written as
\begin{eqnarray} \label{4.14}
K_g & = & \frac{c}{6(1-c)}\Big[\frac{2+c}{9c}H^2 D_1(Z_1\cdot K) \nn\\ &+& H^2x(Z_1\cdot K) - (Z_1\cdot\bar\partial) K \nn\\ &+& \frac{2-5c}{c}H^2(x\cdot Z_1)K\Big] + \Lambda_g,\;\; c\neq 1.
\end{eqnarray}
Interestingly, we have $ x\cdot K_g=0,\;\; \bar\partial\cdot K_g= \frac{1}{3}H^2 Z_1\cdot K$, which verify Eq. (\ref{4.2}). Note that $\Lambda_g$ is a vector field:
$$(Q_1+6)\Lambda_g=0,\;\; x\cdot\Lambda_g=0,\;\; \bar\partial\cdot\Lambda_g=0.$$
Actually, it is the gauge solution characterized in the previous section.

Accordingly, using Eqs. (\ref{4.8}), (\ref{4.9}), and (\ref{4.14}), the tensor field ${\cal{K}}^t$ can be written in the following form,
\begin{equation} \label{4.18} {\cal{K}}^t_{\alpha \beta}(x)={\cal D}_{\alpha \beta}(x,\partial,Z_1,Z_2)\phi_2,\end{equation}
where ${\cal D}$ is the projector tensor ($c\neq1$):
\begin{eqnarray}
{\cal D}(x,\partial,Z_1,Z_2) = \hspace{6cm}\nn\\
\Big( -\frac{2}{3} \theta Z_1\cdot +{\cal S}\bar Z_1 + \frac{c}{6(1-c)}D_2 \Big[ \frac{2+c}{9c}H^2 D_1(Z_1\cdot )\hspace{1cm} \nn\\
+ H^2x(Z_1\cdot ) - (Z_1\cdot\bar\partial) + \frac{2-5c}{c}H^2(x\cdot Z_1) \Big]\Big)\hspace{0.6cm} \nn\\
\times \Big( \bar Z_{2} - \frac{1}{2}D_1[(Z_2\cdot \bar\partial) + 2H^2(x\cdot Z_2)] \Big).\hspace{0.5cm}
\end{eqnarray}

\subsection{Plane wave method}
An axiomatic field theory in dS space based on analyticity in the complexified Riemannian manifold has been developed by Bros, Gazeau, and Moschella through the extension of the Fourier-Helgason transformation in dS space (see \cite{bros1,bros2} and references therein). They found coordinate-independent plane waves in dS space which are eigenfunctions of the Laplace-Beltrami operator relative to the geometry of the curved space and play the role of the plane waves in Minkowski space.

Here, by exploiting the plane wave formalism, we explicitly show that applying the extra condition, $\bar\partial\cdot K= 0$, automatically removes the logarithmic divergence terms from the equations.

In the form of the dS plane wave, $\phi_2\equiv\phi$, the "massless" minimally coupled scalar field is given by \cite{bros1,bros2}
\begin{equation}\label{A.1} \phi = (Hx\cdot\xi)^\sigma,\;\;\; \sigma=0,-3\end{equation}
where this 5-vector $\xi$ lies on the positive null cone ${\cal C}^{+} = \{ \xi \in \Re^5;\;\;\xi^2=0,\; {\xi}^0>0 \}$.

In this respect, it is the work of a few lines to write Eq. (\ref{4.18}) in the following form (see Appendix C):
$${\cal{K}}^t_{\alpha \beta}(x)={\cal{E}}_{\alpha \beta}(x,\xi,Z_1,Z_2)(Hx\cdot\xi)^\sigma,\;\;\; \sigma=0,-3$$
in which ($c\neq1$),
\begin{eqnarray}\label{A.7}
{\cal{E}}_{\alpha\beta}={\cal{S}}\Big[{{\cal{E}}_1(c,\sigma)}\bar{Z}_{1\alpha}\bar{Z}_{2\beta}+{{\cal{E}}_2(c,\sigma)}\bar{Z}_{1\alpha}\bar{\xi}_{\beta}\hspace{1.5cm}\nn\\
+{{\cal{E}}_3(c,\sigma)}\bar{Z}_{2\alpha}\bar{\xi}_{\beta}+{{\cal{E}}_4(c,\sigma)}\bar{\xi}_\alpha\bar{\xi}_\beta +{{\cal{E}}_5(c,\sigma)}\theta_{\alpha\beta} \Big],
\end{eqnarray}
where
$${{\cal{E}}_1(c,\sigma)}=-\frac{\sigma}{2}+\frac{\sigma c}{12(c-1)}\Big[ 2(\sigma+3)\frac{2+c}{9c}+\frac{2-5c}{c}-(2\sigma+3) \Big],$$
$${{\cal{E}}_2(c,\sigma)}=\Big(-\frac{1}{2}\sigma(\sigma+2)+\frac{\sigma c}{12(c-1)}\Big[ 2\sigma(\sigma+3)\frac{2+c}{9c}$$ $$+(\sigma+2)\frac{2-5c}{c}-\sigma(2\sigma+5) \Big]\Big)\frac{x\cdot Z_2}{x\cdot\xi},$$
$${{\cal{E}}_3(c,\sigma)}=\frac{\sigma c}{6(c-1)}\Big[ \sigma(\sigma+3)\frac{2+c}{9c}+(\sigma+2)\frac{2-5c}{c}$$ $$-\sigma(\sigma+1) \Big]\frac{x\cdot Z_1}{x\cdot\xi},$$
$${{\cal{E}}_4(c,\sigma)}=\frac{\sigma(\sigma-1)c}{12(c-1)}\Big[ H^{-2}\Big( \sigma\frac{2+c}{9c}-2-\sigma \Big)\frac{Z_1\cdot Z_2}{(x\cdot\xi)^2}$$ $$+\Big( \sigma(\sigma+3)\frac{2+c}{9c}+(\sigma+2)\frac{2-5c}{c}$$ $$-\sigma(\sigma+2) \Big)\frac{(x\cdot Z_1)(x\cdot Z_2)}{(x\cdot\xi)^2} \Big], $$
\vspace{4mm}$${{\cal{E}}_5(c,\sigma)}=\frac{\sigma}{6} \Big[ \Big( 1+\frac{c}{2(c-1)}\Big[ \sigma\frac{2+c}{9c}-\sigma-2 \Big] \Big)Z_1\cdot Z_2$$
$$+(\sigma+3)\Big( 1+\frac{1}{9(c-1)}(11-26c+\sigma -4c\sigma)H^2(x\cdot Z_1)(x\cdot Z_2) \Big) \Big].$$
One can also easily show
\begin{eqnarray}\label{A.8}
({\cal{K}}^t)'=\frac{\sigma(\sigma+3)}{27(c-1)}\Big[ {{\cal{E}}'_1(c,\sigma)}(Z_1\cdot Z_2)\hspace{2cm}\nn\\
+{{\cal{E}}'_2(c,\sigma)}H^2(x\cdot Z_1)(x\cdot Z_2) \Big]\phi,
\end{eqnarray}
where
$${{\cal{E}}'_1(c,\sigma)}=2+\sigma-4c\sigma-17c,$$
$${{\cal{E}}'_2(c,\sigma)}=46-121c+2(8-23c)\sigma+(1-4c)\sigma^2.$$
In our case $\sigma=0,-3$, we have $({\cal{K}}^t)'=0$.

Obviously, by putting $\sigma=0,-3$, no logarithmic singular term appears in the solution. Therefore, the gauge-fixing parameter $c$ does not need to be fixed to the value of $2/5$. Nonetheless, we must emphasize that the gauge-fixing is an unavoidable procedure for quantization of the tensor field. In this regard see section VII.

\section{The Krein-Gupta-Bleuler structure lying behind the dS minimally coupled massless field}
In the previous section, we showed that on a dS background, the traceless part (the spin-two sector) of tensor field ${\cal{K}}$ can be written in terms of the massless minimally coupled scalar field. Covariant quantization of this field, therefore, would be interestingly important in building the dS quantum linear gravity. Here, we review the Krein-Gupta-Bleuler formalism proposed in \cite{de Bievre6230,Gazeau1415} which yields a fully covariant quantization of the massless minimally coupled scalar field.

Through extensive investigations of the quantization of this field \cite{Allen3136,Allen3771,Polarski1892,Bertola,Kristen567,Chernikov109}, it turns out that in obtaining a covariant construction of the propagator function for the field, one encounters the difficulty that the Laplace-Beltrami operator $\Box_H$ has a normalizable zero-frequency mode (more precisely a constant mode) on the Euclidean continuation of dS, $S^4$. As a result, no dS-invariant propagator inverse for the wave operator $\Box_H$ exists. Indeed, the infrared divergence appears. It should be mentioned that this result is not an artifact of the Euclidean continuation since it has been shown by Allen \cite{Allen3136} that there exists no Hilbertian-de Sitter covariant Fock vacuum for the massless minimally coupled field. To have a deeper insight into this difficulty, we notice that the zero-frequency mode has a positive norm, but it is not part of the Hilbertian structure of the one-particle sector. Indeed, regarding the conformal time, all the negative frequency solutions of the field equation are generated by applying the de Sitter group action on this mode. Consequently, the Hilbertian Fock space (built of any complete set of modes including the zero mode; ${\cal{H}}_{+}=\{\sum_{k\geq0} \alpha_k\phi_k; \sum_{k\geq0}|\alpha_k|^2<\infty \}$, $\phi_k$ is defined in \cite{Gazeau1415}) is not de Sitter-invariant or, more precisely, is not closed under the de Sitter group action. Actually, the critical point about the minimally coupled field is originated in the nonexistence of a covariant decomposition, ${\cal{H}}_{+} \oplus {\cal{H}}_{-}$,\footnote{${\cal{H}}_{-}$ is the corresponding anti-Hilbert space, a space with definite negative inner product.} (none of ${\cal{H}}_{+}$ and ${\cal{H}}_{-}$ carry a dS group representation). Note that, there exists such a decomposition for the massive scalar field case, in which the usual space of physical states is ${\cal{H}}_{+}$ verifying ${\cal{H}_{+}^*}={\cal{H}}_{-}$ \cite{Gazeau1415,de Bievre6230}.

Interestingly, there is a profound analogy between this difficulty and the quantum electrodynamic case. Considering a constant function $\lambda$, there exists a gaugelike global transformation for the Lagrangian,
$${\cal{L}}=\sqrt{|g|}\partial_\mu \phi \partial^\mu \phi,$$
of the free field. It is invariant under $\phi\rightarrow\phi + \lambda$. So, it would not be surprising if a generalization of the Gupta-Bleuler procedure would serve identically for this situation. Indeed, the representation structure of the minimally coupled scalar field requires another Gupta-Bleuler type of triplet (not the one proposed in Sec. III to deal with the gauge invariance of the Eq. (\ref{2.26})), where the gauge states are the constant functions \cite{Gazeau1415,de Bievre6230}. An appropriate adaptation (Krein spaces) of the Wightman-G\"{a}rding axiomatic for massless fields (Gupta-Bleuler scheme) \cite{Wightman} fulfils this requirement, so that, the space of gauge states is simply materialized as the space of constant functions $\cal{N}$ in the "one-particle sector" of the field which plays the role of $V_g$ (see Sec. III). While, the physical one-particle space, called $\cal{M}$, is a space of positive frequency solutions of the field equation where the Klein-Gordon inner product is positive but degenerate. Note that, the covariance of the field would be broken by exploiting the canonical quantization structured into a degenerate space of solutions. Accordingly, one must construct a larger space ${\cal{H}}$, called the total space, which is a nondegenerate invariant space of solutions and admits $\cal{M}$ as an invariant subspace. These spaces are ingredients of the Krein-Gupta-Bleuler triplet ${\cal{N}}\subset{\cal{M}}\subset {\cal{H}}$; it is proved that ${\cal{H}}$ is a Krein space, i.e. ${\cal{H}}={\cal{H}}_{+} \oplus {\cal{H}}_{-}$ \cite{Gazeau1415,de Bievre6230}.

As usual in a Gupta-Bleuler model, the quantum field is written rigorously as an operator-valued distribution on a Fock space built on ${\cal{H}}$, in which the Klein-Gordon inner product is nondegenerate, but not positively definite. Because of the appearance of the negative norm states, however, the total space ${\cal{H}}$ cannot be considered as the physical states space. Therefore, in order to have a reasonable interpretation of the theory guaranteed, the selection of the subspace of physical states is required. In stationary spacetimes (Minkowski) case, there exists a Killing vector field $X$ which is timelike at each point, and therefore one can define a Hamiltonian for the quantization space: $iX$. Accordingly, by admitting a positive spectrum for the Hamiltonian, the physical states space is determined. This procedure for dS spacetime which is not stationary, and so there is no timelike Killing vector for it, can be performed by demanding that the physical states be positive frequencies regarding the conformal time on dS spacetime. It has been shown that the quotient space ${\cal{M}} / {\cal{N}}$, a Hilbert space carrying the UIR of the de Sitter group, characterizes the set of physical states \emph{sensu stricto}. It turns out that the theory which has been resulted through this procedure has all the properties one might require from a free field on a spacetime with high symmetry \cite{Gazeau1415,de Bievre6230}.

It is proved that the above construction provides a causal and fully covariant quantum field of the dS minimally coupled massless scalar field which is also free of infrared divergence \cite{Gazeau1415,de Bievre6230}. We must underline that, here, there is no contradiction with Allen's theorem cited above; our considered field is constructed over a non-Hilbertian Fock space. More accurately, the one-particle sector $\cal{M}$ itself is not a Hilbert space (the inner product is positive but degenerate).

Technically as mentioned, through the Krein construction, the field acts on a states space of the Fock space structure but including both positive and negative norm states
\begin{equation} \phi(x)=\frac{1}{\sqrt{2}}[\phi_{+}(x)+\phi_{-}(x)],\end{equation}
in which
\begin{eqnarray}\phi_{+}(x)&=&\sum_{k\geq0}(a_k\phi_k(x) + a^\dag_k\phi^\ast_k(x)),\nn\\
\phi_{-}(x)&=&\sum_{k\geq0}(b^\dag_k\phi_k(x) + b_k\phi^\ast_k(x)).\end{eqnarray}
Here, the positive mode $\phi_{+}(x)$ is the scalar field that was used by Allen \cite{Allen3136,Allen3771}. A significant difference between this canonical quantization approach and the standard QFT, which is based on canonical commutation relations, lies in the requirement of the following commutation relations,
\begin{equation}\label{ccr}
[a_k,a^\dagger_{k'}]=\delta_{kk'},\;\;\; [b_k,b^\dagger_{k'}]=-\delta_{kk'}.
\end{equation}
The other commutation relations are zero.

The (Krein-)Fock vacuum, $|\Omega\rangle$, is specified by
\begin{equation}\label{Fockvacuum}
a_k|\Omega\rangle=0,\;\;\; b_k|\Omega\rangle=0.
\end{equation}
It is invariant under the dS group action \cite{Gazeau1415}. More accurately, in the Krein context, the Fock vacuum is unique and normalizable. It is independent of the Bogolubov transformations \cite{Gazeau1415}. This does not, however, concern us since in this construction not only is the vacuum different but so is the field itself. The point is indeed within the concept of how to determine an observable in the Gupta-Bleuler formalism. Actually, defining observables is performed through the feature that they do not "see" the gauge states. As a result, the field itself is not an observable (it is gauge dependent). Nevertheless, regarding the fact that $\mu$ refers to global coordinates, the physically interesting observables (e.g. the energy-momentum tensor) can be built using operators $\partial_\mu$ on the total space \cite{de Bievre6230}. While, as usual in a Gupta-Bleuler construction, the average values of observables will be evaluated only with physical states,
$$|\vec{k}\rangle\equiv|{k}_1^{n_1}...{k}_l^{n_l}\rangle = \frac{1}{\sqrt{n_1!...n_l!}}(a_{k_1}^\dag)^{n_1}...(a_{k_l}^\dag)^{n_l}|\Omega\rangle,$$
which are in fact the elements of ${\cal{M}} / {\cal{N}}$.

In this construction, therefore, the fact that $\phi$ is not an observable implies that the different two-point functions, like Wightman or Hadamard functions,
$$\langle\Omega|\phi(x)\phi(x') |\Omega\rangle, \;\;\langle\Omega|\phi(x)\phi(x') + \phi(x')\phi(x)|\Omega\rangle,$$
are gauge dependent. As an example, the symmetric two-point function (Hadamard function) is not expected to have great meaning in our construction, and a straightforward computation indeed shows that it vanishes. The crucial point is that any definition \emph{a priori} of such a function cannot yield a covariant theory; there exists no nontrivial covariant two-point function of the positive type for the minimally coupled quantum field on de Sitter spacetime \cite{Allen3136,Allen3771,Bertola}. Therefore, this result is nothing but another formulation of Allen's theorem cited above. Indeed, the only two-point function which naturally appears is the commutator, but it is not of the positive type and it does not allow us to select physical states \cite{Gazeau1415,de Bievre6230}. In addition, one should pay attention that in this construction, the link between the vacuum and the two-point function is not the same as the standard QFT. The standard classification of vacua is based on two-point functions, and the Krein vacuum does not fit this classification. In this context, contrary to the usual QFT for which to choose a vacuum is to choose a physical space of states and a two-point function, the vacuum is unique and does not characterize the physical space of states.

Here, we must underline that the invariance of the Fock vacuum does not imply that the Bogolioubov transformations, which merely modify the set of physical states, are no longer valid in this quantization method. Admittedly, not only the selected spacetime but also the observer affect the space of physical states; an accelerated observer in Minkowski space has a different set of the physical states from those of an inertial observer (Unruh effect), while both observers have the same field representation. Indeed, in this context, "instead of having a multiplicity of vacua, we have several possibilities for the space of physical states and only one field and one vacuum which are independent of Bogolubov transformations. More precisely, the usual ambiguity about vacua is not suppressed but displaced " \cite{Garidi}.

A direct consequence of this construction is an automatic covariant renormalization of the energy-momentum tensor. Actually, a trivial computation of the mean values of the components of the energy-momentum tensor reveals that
$$|\langle k_1^{n_1}... k_l^{n_l}|T_{\mu\nu}| k_1^{n_1}... k_l^{n_l}\rangle|<\infty,$$
and in spite of the presence of negative norm modes in the theory, no negative energy can be measured for any physical state,
$$\langle k_1^{n_1}... k_l^{n_l}|T_{00}| k_1^{n_1}... k_l^{n_l}\rangle\geq0.$$
This quantity vanishes if and only if $|\vec{k}\rangle=|\Omega\rangle$. Interestingly, it is proved that this renormalizing procedure completely fulfils the so-called Wald axioms \cite{Gazeau1415}. Moreover, it is worth mentioning that due to the vanishing of the vacuum expectation value of the energy-momentum tensor $\langle \Omega|T_{\mu\nu}|\Omega\rangle=0$, the so-called conformal anomaly disappears from the trace of the energy-momentum tensor, while, all other renormalization methods present this anomaly. So, from this perspective, it seems that we face a very different renormalization scheme. However it is not very surprising, since our construction preserves covariance and conformally covariance in a rather strong sense \cite{de Bievre6230}. As a result, the model does not exhibit the trace anomaly which, after all, can appear only by breaking the conformal invariance.

The behavior of the method in Minkowski spacetime, especially when interaction is present, has also been investigated in Ref. \cite{Garidi}. In this regard, it is proved that thanks to the condition that preserves the unitarity of the theory (see Appendix D), the method is capable of retrieving the results of (Hilbert space) QFT's counterpart with the exception that the free-field vacuum energy vanishes, without any reordering nor regularization.

Moreover, the Krein space quantization approach to Hawking radiation has been investigated in Ref. \cite{Pejhan}. It is well known that the study of quantum field theories on a gravitational background culminates in the celebrated Hawking prediction of black hole evaporation \cite{Hawking199}. Actually, Hawking radiation is of great significance since it unites gravitational physics near strong gravitational objects like black holes and quantum field theory. Therefore, it can be considered a milestone of modern theoretical physics which provides a test bed for candidate theories of Quantum Gravity. Accordingly, any consistent quantum field theoretical approach to gravity must include Hawking radiation. In this respect, in Ref. \cite{Pejhan}, by proposing a model to simulate schwarzschild black holes, it is shown that by utilizing the Krein space quantization one obtains the very result for Hawking radiation. In this regard, see also \cite{SH1408}.

In the following section, respecting the above capabilities, the Krein-Gupta-Bleuler construction is considered to calculate the two-point function of the spin-two part of the linear quantum gravity in dS space.

\section{The two-point function (the spin-two sector)}
The associated Wightman two-point function for the spin-two part of the linear quantum gravity in dS space is dealt with in this section. In this regard, the two-point function is written in terms of bitensors (these are functions of two points $(x, x')$ and at each point behave like tensors under coordinate transformations \cite{allen2}). If bitensors preserve the dS invariance, we call them maximally symmetric. Bitensor Wightman two-point functions are the cornerstone of the dS axiomatic field theory construction \cite{bros2}.

On this basis, the two-point function is delivered by
\begin{eqnarray} \label{6.1}
{\cal{W}}^t_{\alpha\beta\alpha'\beta'}(x,x')= \langle\Omega|{\cal{K}}^t_{\alpha\beta}(x){\cal{K}}^t_{\alpha'\beta'}(x')|\Omega\rangle, \nn\\ \alpha,\beta=0,1,2,3,4.
\end{eqnarray}
where $x,x'\in X_H$ and $|\Omega\rangle $ is the (Krein-)Fock-vacuum state. The two-point function must verify the field equation (\ref{2.26}), with regard to $x$ and $x'$ (without any difference), and the following physical requirements as well;
\begin{itemize}
\item{\textbf{Indefinite sesquilinear form}\\
For any test function $f_{\alpha\beta}\in{\cal{D}}(X_H)$, an indefinite sesquilinear form is defined by
\begin{equation}\label{6.4} \int_{X_H\times X_H}f^{\ast\alpha\beta}(x){\cal{W}}^t_{\alpha\beta\alpha'\beta'}(x,x')f^{\alpha'\beta'}(x')d\sigma (x)d\sigma (x'), \end{equation}
in which $f^\ast$ is the complex conjugate of $f$ and $d\sigma(x)$ determines the de Sitter-invariant measure on $X_H$. ${\cal{D}}(X_H)$ is the space of functions $C^{\infty}$ with compact support in $X_H$.}

\item{\textbf{Locality}\\
For every space-like separated pair $(x,x')$, \textit{i.e.} $x\cdot x'>-H^{-2}$,
\begin{equation}\label{6.5} {\cal{W}}^t_{\alpha\beta\alpha'\beta'}(x,x')={\cal{W}}^t_{\alpha'\beta'\alpha\beta}(x',x). \end{equation}}

\item{\textbf{Covariance}
\begin{equation}\label{6.6} (g^{-1})_\alpha^\gamma(g^{-1})_\beta^\delta{\cal{W}}^t_{\gamma \delta \gamma' \delta'}(gx,gx')g_{\alpha'}^{\gamma'}g_{\beta'}^{\delta'}={\cal{W}}^t_{\alpha\beta\alpha'\beta'}(x,x'), \end{equation}
for all $g\in SO_0(1,4)$.}

\item{\textbf{Index symmetrizer}
\begin{equation}\label{6.7} {\cal{W}}^t_{\alpha\beta\alpha'\beta'}(x,x')={\cal{W}}^t_{\beta\alpha\beta'\alpha'}(x,x'). \end{equation}}

\item{\textbf{Transversality}
\begin{equation}\label{6.8} x^\alpha{\cal{W}}^t_{\alpha\beta\alpha'\beta'}(x,x')=0=x'^{\alpha'}{\cal{W}}^t_{\alpha\beta\alpha'\beta'}(x,x'). \end{equation}}

\item{\textbf{Tracelessness}
\begin{equation}\label{6.9} ({\cal{W}}^t)_{\;\;\;\alpha\alpha'\beta'}^\alpha(x,x')=0={({\cal{W}}^t)_{\alpha\beta\alpha'}}^{\alpha'}(x,x'). \end{equation}}
\end{itemize}

In this respect, by considering Eqs. (\ref{4.1}) and (\ref{6.1}), the most general dS-invariant form for a transverse two-point function can be written as
\begin{eqnarray}\label{6.10} {\cal{W}}^t(x,x')=\theta \theta' {\cal{W}}_0(x,x')+{\cal{S}}{\cal{S}}'\theta\cdot\theta'{\cal{W}}_1(x,x')\nn\\
+D_2D'_2{\cal{W}}_g(x,x'),\;\;\;\;\;\;\;\;\;\;\;\;\;\;\;\;\;\;\;\;\;\end{eqnarray}
where ${\cal W}_1$ and ${\cal W}_g$ are transverse bi-vectors, ${\cal W}_0$ is bi-scalar and $D_2D'_2= D'_2D_2$.

With regard to the above considerations, we choose $x$ to start investigation. The two-point function (\ref{6.10}) must satisfy Eq. (\ref{2.26}), in this respect, it could be easily shown\footnote{Note that, the primed operators act on the primed coordinate only and vise versa.}
\begin{equation}\label{6.11}
\left\{\begin{array}{rl} &(Q_0+6)\theta'{\cal W}_0=-4{\cal S}'\theta'\cdot {\cal W}_{1},\hspace{1.7cm}(I)\vspace{2mm}\\
&(Q_1+2){\cal W}_{1}=0, \hspace{3.5cm}(II)\vspace{2mm}\\
&(Q_1+6)D'_2{\cal W}_g= \frac{c}{2(c-1)} H^2 D_1 \theta'{\cal W}_0\\
&\hspace{2.5cm}+ H^2{\cal S}'\Big[\frac{2-5c}{1-c}(x\cdot\theta')\\
&\hspace{2.5cm}+ \frac{c}{1-c}(D_1\theta'\cdot - x\theta'\cdot \\
&\hspace{2.5cm}- H^{-2}\theta'\cdot\bar\partial)\Big]{\cal W}_{1}.\hspace{1cm}(III)
\end{array}\right.
\end{equation}
where the condition $\partial\cdot{\cal W}_1 = 0$, is applied. Considering Eqs. (\ref{6.11}-I) and (\ref{6.11}-II), it yields
\begin{equation} \label{6.12} \theta'{\cal W}_0(x,x')=-\frac{2}{3}{\cal S}'\theta'\cdot{\cal W}_{1}(x,x').\end{equation}
The bi-vector two-point function ${\cal W}_{1}$, which is the solution of Eq. (\ref{6.11}-II), can be written as
$${\cal W}_{1}=\theta\cdot\theta'{\cal W}_{2}+D_1D'_1{\cal W}_{3}.$$
where ${\cal W}_{2}$ and ${\cal W}_{3}$ are bi-scalar two-point functions, so that
$$D'_1{\cal W}_{3}=-\frac{1}{2}[2H^2 (x\cdot\theta'){\cal W}_2 + \theta'\cdot\bar\partial{\cal W}_{2}],$$
$$Q_0{\cal W}_{2}=0.$$
Therefore, ${\cal W}_{2}\equiv{\cal W}_{mc}$ is a massless minimally coupled bi-scalar two-point function. With regard to the above identities, the bi-vector two-point function would be
\begin{equation}\label{6.13} {\cal W}_{1}(x,x')=\Big(\theta\cdot\theta' - \frac{1}{2}D_{1}[\theta'\cdot\bar\partial + 2H^2 x\cdot\theta']\Big){\cal W}_{mc}(x,x').\end{equation}

Pursuing a similar procedure utilized in Appendix B, one obtains
$$ (Q_1+6) x\theta' \cdot {\cal W}_{1} = 6 x\theta' \cdot {\cal W}_{1},$$
$$ (Q_1+6)D_1 \theta'\cdot{\cal W}_{1} = 6 D_1 \theta'\cdot{\cal W}_{1}, $$
$$ (Q_1+6)\theta'\cdot\bar\partial{\cal W}_{1} = 6 \theta'\cdot\bar\partial{\cal W}_{1} + 2H^2 D_1 (\theta'\cdot{\cal W}_{1}), $$
$$ (Q_1+6)^{-1}(x\cdot\theta'){\cal W}_{1} = \frac{1}{6}\Big[\frac{1}{9} D_1 (\theta'\cdot{\cal W}_{1}) + (x\cdot\theta'){\cal W}_{1}\Big].$$
Using the aforementioned identities along with Eqs. (\ref{6.11}-III) and (\ref{6.12}), we obtain
\begin{eqnarray}\label{6.14}
D'_2 {\cal W}_{g}(x,x') & = & \frac{cH^2}{6(1-c)}{\cal S}' \Big[ \frac{2+c}{9c}D_1 \theta'\cdot{\cal W}_{1} \nn\\ &+& \frac{2-5c}{c} x\cdot\theta'{\cal W}_{1} + x\theta'\cdot{\cal W}_{1} \nn\\ &-& H^{-2} \theta'\cdot\bar\partial{\cal W}_{1} \Big], \;\; c\neq1.
\end{eqnarray}

Correspondingly, the two-point function (\ref{6.10}) would be
\begin{equation} \label{6.18'} {\cal W}^t_{\alpha\beta\alpha'\beta'}(x,x')= \Delta_{\alpha\beta\alpha'\beta'}(x,x'){\cal W}_{mc}(x,x'), \end{equation}
where ($c\neq1$)
\begin{eqnarray}\label{6.19'}
\Delta (x,x')=& -\frac{2}{3}{\cal S}'\theta\theta'\cdot \Big(\theta\cdot\theta' - \frac{1}{2}D_{1}[\theta'\cdot\bar\partial + 2H^2 x\cdot\theta']\Big)\nn\\
&+{\cal S}{\cal S}'\theta\cdot\theta'\Big(\theta\cdot\theta' - \frac{1}{2}D_{1}[\theta'\cdot\bar\partial + 2H^2 x\cdot\theta']\Big)\nn\\
&+\frac{cH^2}{6(1-c)}{\cal S}'D_2 \Big(\frac{2+c}{9c}D_1 \theta'\cdot + \frac{2-5c}{c} x\cdot\theta' + x\theta'\cdot \nn\\
&\hspace{3cm} - H^{-2} \theta'\cdot\bar\partial\Big)\nn\\
&\times\Big(\theta\cdot\theta' - \frac{1}{2}D_{1}[\theta'\cdot\bar\partial + 2H^2 x\cdot\theta']\Big).
\end{eqnarray}

On the other hand, the two-point function (\ref{6.10}) must verify Eq. (\ref{2.26}) with regard to $x'$. So, pursuing the same procedure, we have
$$ \left\{\begin{array}{rl} &(Q'_0+6)\theta{\cal W}_0=-4{\cal S}\theta\cdot{\cal W}_{1},\hspace{1.9cm}(I)\vspace{2mm}\\
&(Q'_1+2){\cal W}_{1}=0, \hspace{3.5cm}(II)\vspace{2mm}\\
&(Q'_1+6)D_2{\cal W}_g= \frac{c}{2(c-1)} H^2 D'_1 \theta{\cal W}_0\\
&\hspace{2.5cm}+ H^2{\cal S}\Big[\frac{2-5c}{1-c}(x'\cdot\theta)\\
&\hspace{2.5cm}+ \frac{c}{1-c}(D'_1\theta\cdot - x'\theta\cdot\\
&\hspace{2.5cm}- H^{-2}\theta\cdot\bar\partial')\Big]{\cal W}_{1}.\hspace{1cm}(III)
\end{array}\right.$$
here, the condition $\partial'\cdot{\cal W}_1 = 0$ is implemented. In this case, we have
\begin{equation} \label{6.15} \theta{\cal W}_0(x,x')= -\frac{2}{3}{\cal S}\theta\cdot{\cal W}_{1}(x,x'),\end{equation}
\begin{equation}\label{6.16} {\cal W}_{1}(x,x')=\Big(\theta\cdot\theta' - \frac{1}{2}D'_{1}[\theta\cdot\bar\partial' + 2H^2 x'\cdot\theta]\Big){\cal W}_{mc}(x,x'),\end{equation}

\begin{eqnarray}\label{6.17}
D_2 {\cal W}_{g}(x,x') & = & \frac{cH^2}{6(1-c)}{\cal S}\Big[\frac{2+c}{9c}D'_1 \theta\cdot{\cal W}_{1} \nn\\ &+& \frac{2-5c}{c} x'\cdot\theta{\cal W}_{1} + x'\theta\cdot{\cal W}_{1} \nn\\ &-& H^{-2} \theta\cdot\bar\partial'{\cal W}_{1} \Big], \;\; c\neq1.
\end{eqnarray}

Utilizing Eqs. (\ref{6.15})-(\ref{6.17}) it turns out that the bitensor two-point function can be written in the following form
\begin{equation} \label{6.18} {\cal W}^t_{\alpha\beta\alpha'\beta'}(x,x')= \Delta'_{\alpha\beta\alpha'\beta'}(x,x'){\cal W}_{mc}(x,x'), \end{equation}
where ($c\neq1$)
\begin{eqnarray}\label{6.19}
\Delta' (x,x')=& -\frac{2}{3}{\cal S}'\theta'\theta\cdot \Big(\theta'\cdot\theta - \frac{1}{2}D'_{1}[\theta\cdot\bar\partial' + 2H^2 x'\cdot\theta]\Big)\nn\\
&+{\cal S}{\cal S}'\theta'\cdot\theta\Big(\theta'\cdot\theta - \frac{1}{2}D'_{1}[\theta\cdot\bar\partial' + 2H^2 x'\cdot\theta]\Big)\nn\\
&+\frac{cH^2}{6(1-c)}{\cal S}D'_2 \Big(\frac{2+c}{9c}D'_1 \theta\cdot + \frac{2-5c}{c} x'\cdot\theta + x'\theta\cdot \nn\\
&\hspace{3cm} - H^{-2} \theta\cdot\bar\partial'\Big)\nn\\
&\times\Big(\theta'\cdot\theta - \frac{1}{2}D'_{1}[\theta\cdot\bar\partial' + 2H^2 x'\cdot\theta]\Big).
\end{eqnarray}

In summary, thus far by using an ansatz analogous to the one used for calculating the field solutions, we have shown that the spin-two part of the graviton two-point function can be written in terms of the scalar massless minimally coupled two-point function ${\cal W}_{mc}$. As already discussed in Sec. V, to obtain a de Sitter fully covariant construction for the minimally couple massless scalar field, the Krein-Gupta-Bleuler quantization formalism should be in order; there is no nontrivial covariant two-point function of positive type \cite{Allen3136,Allen3771}, and the only two-point function which naturally appears is the commutator, but it is not of positive type \cite{Gazeau1415,de Bievre6230}. In this respect, if one requires the function ${\cal W}_{mc}$ to be de Sitter-invariant (and ignores its analyticity properties for the time being \cite{Bertola}), it will only depend on the invariant length ${\cal{Z}}\equiv -H^2x\cdot x'$; ${\cal W}_{mc} = {\cal W}_{mc}({\cal Z})$. Note that, ${\cal{Z}}(x,x')$ is an invariant object under the isometry group $O(1,4)$ and hence any function of ${\cal Z}$ is dS invariant as well. Accordingly, the equation $Q_0 {\cal W}_{mc}({\cal Z})=0$ becomes the ordinary differential equation (see (\ref{A.16}))
\begin{eqnarray}\label{d/dz}
\Big((1-{\cal Z}^2)\frac{d^2 }{d {\cal Z}^2} - 4 {\cal Z}\frac{d }{d {\cal Z}}\Big) {\cal W}_{mc}({\cal Z}) = 0.
\end{eqnarray}

Now, considering that ${\cal W}_{mc}$ is only a function of ${\cal{Z}}(x,x')$, we can use the identities given in Appendix A to obtain the following expressions:
\begin{widetext}
\begin{equation} \label{6.22} \theta'_{\alpha'\beta'}{\cal W}_{0}(x,x')=  \frac{1}{3}{\cal S}' \Big[\theta'_{\alpha'\beta'} + \frac{4}{1-{\cal{Z}}^2}H^2 (x\cdot\theta'_{\alpha'})(x\cdot\theta'_{\beta'})\Big]{\cal{Z}}\frac{d}{d{\cal{Z}}}{\cal W}_{mc}({\cal{Z}}),
\end{equation}
\begin{equation} \label{6.23}
{\cal W}_{1\beta\beta'}(x,x')=  \frac{1}{2} \Big[\frac{3+{\cal{Z}}^2}{1-{\cal{Z}}^2}H^2 (x'\cdot\theta_{\beta})(x\cdot\theta'_{\beta'}) - {\cal{Z}} (\theta_{\beta}\cdot\theta'_{\beta'})\Big]\frac{d}{d{\cal{Z}}}{\cal W}_{mc}({\cal{Z}}),
\end{equation}
\begin{eqnarray}\label{6.24}
D_{2\alpha}D'_{2\alpha'}{\cal W}_{g\beta\beta'}(x,x') & = & -\frac{H^2}{54(1-c)(1-{\cal{Z}}^2)^2}{\cal S}{\cal S}'\Big[ H^{-2}{\cal{Z}}(1-{\cal{Z}}^2)\Big( 1-13c+3(1-c){\cal{Z}}^2 \Big) \theta_{\alpha\beta} \theta'_{\alpha'\beta'} \nn\\ & + & H^{-2}{\cal{Z}}(1-{\cal{Z}}^2)\Big( 17-41c-9(1-c){\cal{Z}}^2 \Big)(\theta_{\alpha}\cdot\theta'_{\alpha'}) (\theta_{\beta}\cdot\theta'_{\beta'}) \nn\\ & + & 24{\cal{Z}}\Big( 2-5c-(1-c){\cal{Z}}^2 \Big)\theta_{\alpha\beta}(x\cdot\theta'_{\alpha'})(x\cdot\theta'_{\beta'}) \nn\\ & + & 12{\cal{Z}}\Big( 1-7c+(1-c){\cal{Z}}^2 \Big)\theta'_{\alpha'\beta'}(x'\cdot\theta_{\alpha})(x'\cdot\theta_{\beta}) \nn\\ & + & \Big( -79+199c+(-62+230c){\cal{Z}}^2+45(1-c){\cal{Z}}^4 \Big)(\theta_{\alpha}\cdot\theta'_{\alpha'}) (x\cdot\theta'_{\beta'}) (x'\cdot\theta_{\beta}) \nn\\ & + & \frac{12{\cal{Z}}H^2}{1-{\cal{Z}}^2} \Big( 21-57c-2(1+5c){\cal{Z}}^2-3(1-c){\cal{Z}}^4 \Big) (x'\cdot\theta_{\alpha})(x'\cdot\theta_{\beta})(x\cdot\theta'_{\alpha'})(x\cdot\theta'_{\beta'}) \Big]\nn\\
& \times & \frac{d}{d{\cal{Z}}}{\cal W}_{mc}({\cal{Z}}).
\end{eqnarray}
\end{widetext}
By substitution of Eqs. (\ref{6.22})-(\ref{6.24}) into (\ref{6.10}) we obtain the explicit form of the two-point function (actually the Krein two-point function, which as already discussed, is the commutator) in the ambient formalism as follows\footnote{\textbf{Note:} In these calculations, it is assumed that the two points, $x$ and $x'$, are not on the light cone of one another so that $1-{\cal Z}\neq 0$.}
\begin{eqnarray}\label{6.25}
& & {\cal W}^t_{\alpha\beta \alpha'\beta'}(x,x') = {\frac{2{\cal{Z}}}{27(1-c)(1-{\cal{Z}}^2)^2}}{\cal S}{\cal S}' \hspace{1.5cm} \nn\\
\times & \Big[ &\theta_{\alpha\beta}\theta'_{\alpha'\beta'}f_1(c,{\cal{Z}}) + (\theta_{\alpha}\cdot\theta'_{\alpha'})(\theta_{\beta}\cdot\theta'_{\beta'})f_2(c,{\cal{Z}})\nn\\
& &+  H^2\Big(\theta'_{\alpha'\beta'}(x' \cdot \theta_{\alpha}) (x'\cdot\theta_{\beta})\nn\\
& & \hspace{2.2cm} +\theta_{\alpha\beta}(x\cdot\theta'_{\alpha'})(x\cdot\theta'_{\beta'}) \Big)f_3(c,{\cal{Z}})\nn\\
& &+  H^4\Big( (x'\cdot\theta_{\alpha})(x'\cdot\theta_{\beta})(x\cdot\theta'_{\alpha'})(x\cdot\theta'_{\beta'}) \Big)f_4(c,{\cal{Z}})\nn\\
& &+  (\theta_{\alpha}\cdot\theta'_{\alpha'})(x\cdot\theta'_{\beta'})(x'\cdot\theta_{\beta})f_5(c,{\cal{Z}}) \Big] \frac{d}{d{\cal{Z}}}{\cal W}_{mc}({\cal{Z}}),
\end{eqnarray}
in which
$$f_1(c,{\cal Z})=(1-{\cal Z}^2) [2+c+3(c-1){{\cal{Z}}}^2],$$
$$f_2(c,{\cal{Z}}) = (1-{\cal{Z}}^2) [17c-11 + 9(1-c){{\cal{Z}}}^2],$$
$$ f_3(c,{\cal{Z}}) = 3[7c -1 + (c-1){{\cal{Z}}}^2],$$
$$f_4(c,{\cal{Z}}) = - \frac{3}{(1-{\cal Z}^2)}[ 3(7-19c) - 2(1+5c){{\cal{Z}}}^2 - 3(1-c){{\cal{Z}}}^4],$$
$$ f_5(c,{\cal{Z}})= \frac{1}{{\cal{Z}}} [ 10(4-7c)+ (2-44c){\cal{Z}}^2 - 18(1-c){\cal{Z}}^4 ].$$

In addition, respecting the differential equation (\ref{d/dz}), the function ${\cal W}_{mc}$ (the general solution) would be
\begin{eqnarray}\label{d/dz'}
{\cal W}_{mc}({\cal Z}) = C_1 \Big(\frac{1}{1 + {\cal Z}} - \frac{1}{1 - {\cal Z}} + \ln{\frac{{1 - \cal Z}}{1 + {\cal Z}}} \Big) + C_2,\;\;\;\;
\end{eqnarray}
where the choice of real constants $C_1$ and $C_2$ determines the particular solutions (for instance, if $C_1= H^2/4\pi^2$ then this has the same short-distance behavior as a massless two-point function has in flat space). This function presents problems with locality \cite{Bertola}. However, this does not concern us since in the two-point function (\ref{6.25}), this function enters only via its derivative \cite{Gazeau0},
\begin{eqnarray}\label{d/dz''}
\frac{d}{d{\cal Z}}{\cal W}_{mc}({\cal Z}) = \frac{-4C_1}{({\cal Z}^2 - 1)^2},
\end{eqnarray}
which is a local function. Now, by substituting (\ref{d/dz''}) into (\ref{6.25}), one can easily see that the large-distance growth of the two-point function obviously will not be reflected in the calculated two-point function (\ref{6.25}).

\subsection{The two-point function in the dS intrinsic space notation}
Thus far, the two-point function (\ref{6.25}) has been obtained in the ambient space notation. In order to make comparison with other works \cite{Higuchi3077,Higuchi3005}, one needs to project this two-point function onto the intrinsic space. It was proved that any maximally symmetric bitensor could be expanded in terms of three basic tensors (which form a complete set) \cite{allen2}. The coefficients in this expansion are functions of the geodesic distance $\sigma(x, x')$ and the parallel propagator $g_{\mu\nu'}$,
$$ n_\mu = \nabla_\mu \sigma(x, x')\;\;\;,\;\;\; n_{\mu'} = \nabla_{\mu'} \sigma(x,x'),$$
$$ g_{\mu\nu'}=-c^{-1}({\cal{Z}})\nabla_{\mu}n_{\nu'}+n_\mu n_{\nu'}.$$
For $ {\cal{Z}}=-H^2x\cdot x'$, the geodesic distance can be characterized by
\begin{eqnarray}
\left\{\begin{array}{rl}
     {\cal{Z}}&=\cosh (H\sigma ),  \hbox{if $x$ and $x'$ are time-like separated,} \\
     {\cal{Z}}&=\cos (H\sigma ),  \hbox{if $x$ and $x'$are space-like separated.} \\
\end{array}\right.
\end{eqnarray}
In the de Sitter ambient space formalism, the mentioned fundamental bitensors are given by
$$ \bar{\partial}_\alpha \sigma(x,x')\;\;\;,\;\;\;\bar{\partial}_{\beta'}^{'} \sigma(x,x')\;\;\;,\;\;\;\theta_\alpha .\theta'_{\beta'},$$
these are restricted to the de Sitter hyperboloid by
$$ {\cal{T}}_{\mu\nu'}=\frac{\partial x^\alpha}{\partial X^\mu}\frac{\partial x'^{\beta'}}{\partial X'^{\nu'}}T_{\alpha\beta'}.$$

For $ {\cal{Z}}=\cos(H\sigma)$, we have
$$ n_\mu=\frac{\partial x^\alpha}{\partial X^\mu}\bar{\partial}_\alpha \sigma(x,x')= \frac{\partial x^\alpha}{\partial X^\mu} \frac{H(x' \cdot \theta_\alpha)}{\sqrt{1-{\cal{Z}}^2}},$$
$$ n_{\nu'}=\frac{\partial x'^{\beta'}}{\partial X'^{\nu'}}\bar{\partial}_{\beta'}^{'} \sigma(x,x') =\frac{\partial x'^{\beta'}}{\partial X'^{\nu'}} \frac{H(x\cdot\theta'_{\beta'})}{\sqrt{1-{\cal{Z}}^2}},$$

$$ \nabla_\mu n_{\nu'}=\frac{\partial x^\alpha}{\partial  X^\mu}\frac{\partial x'^{\beta'}}{\partial X'^{\nu'}}\theta^\varrho_\alpha \theta'^{\gamma'}_{\beta'}\bar{\partial}_\varrho\bar{\partial}_{\gamma'}^{'} \sigma(x, x') \hspace{1.8cm}$$
$$=c({\cal{Z}})\Big[n_\mu n_{\nu'}{\cal{Z}}-\frac{\partial x^\alpha}{\partial X^\mu}\frac{\partial x'^{\beta'}}{\partial X'^{\nu'}}\theta_\alpha \cdot\theta'_{\beta'}\Big],$$
where $ c({\cal{Z}})\equiv-\frac{H}{\sqrt{1-{\cal{Z}}^2}}.$

For ${\cal{Z}}=\cosh (H\sigma)$,  $n_\mu$ and $n_{\nu'}$ are multiplied by $i$ and so $c({\cal{Z}})=-\frac{iH}{\sqrt{1-{\cal{Z}}^2}}$. Considering both cases, we have
$$g_{\mu\nu'}+({\cal{Z}}-1)n_\mu n_{\nu'}=\frac{\partial x^\alpha}{\partial X^\mu}\frac{\partial x'^{\beta'}}{\partial X'^{\nu'}}\theta_\alpha \cdot\theta'_{\beta'}.$$
Then the dS ambient two-point function is related to the dS intrinsic counterpart as
$$ Q^t_{\mu\nu\mu'\nu'}= \frac{\partial x^\alpha}{\partial X^\mu} \frac{\partial x^\beta}{\partial X^\nu} \frac{\partial x'^{\alpha'}}{\partial X'^{\mu'}} \frac{\partial x'^{\beta'}}{\partial X'^{\nu'}}{\cal{W}}^t_{\alpha\beta\alpha'\beta'}.$$

Eventually, considering the above identities, we obtain the dS intrinsic two-point function in the following form
\begin{eqnarray}\label{6.26} Q^t_{\mu\nu\mu'\nu'}(X,X') & = & \frac{2{\cal{Z}}}{27(1-c)}{\cal{S}}{\cal{S}}'\Big[  \frac{f_1}{(1-{\cal{Z}}^2)^2}g_{\mu\nu}g'_{\mu'\nu'} \nn\\ & + & \frac{f_2}{(1-{\cal{Z}}^2)^2}g_{\mu\mu'}g'_{\nu\nu'}\nn\\ & + & \frac{f_3}{1-{\cal{Z}}^2}(g_{\mu\nu}n_{\mu'}n_{\nu'}+g'_{\mu'\nu'}n_\mu n_\nu)\nn\\
& + & \Big( \frac{2({\cal{Z}}-1)f_2}{(1-{\cal{Z}}^2)^2}+\frac{f_5}{1-{\cal{Z}}^2} \Big)g_{\mu\mu'}n_\nu n_{\nu'}\nn\\
& + & \Big( \frac{f_2}{(1+{\cal{Z}})^2}-\frac{f_5}{1+{\cal{Z}}}+f_4 \Big)n_\mu n_\nu n_{\mu'}n_{\nu'} \Big]\nn\\
& \times & \frac{d}{d{\cal{Z}}}{\cal{W}}_{mc}({\cal{Z}}).\hspace{0.5cm}
\end{eqnarray}

\section{Considering the spin-zero (pure-trace) part of the theory and the gauge-fixing procedure}
In this section, we study the pure-trace part of ${\cal{K}}$ by considering
$${\cal{K}}^{pt}=\frac{1}{4}\theta \psi,$$
$\psi$ is a scalar field. Taking the trace of the field equation (\ref{2.26}), we have
$$(Q_0+6)\psi+\frac{c}{2}Q_0\psi=0,$$
or equivalently
\begin{equation}\label{66}(Q_0+\frac{12}{c+2})\psi=0,\;\;\;c\neq-2.\end{equation}
It is, however, known that any scalar field in correspondence with the scalar discrete series UIR of the dS group complies the following equation with integer $n$ \cite{Dixmier9}
\begin{equation}\label{77}(Q_0+n(n+3))\psi=0.\end{equation}

In this regard, comparing Eq. (\ref{66}) with (\ref{77}) leads to nontrivial results for the gauge-fixing procedure. As already mentioned, respecting the physical representations of the dS group in the absence of the condition $\bar\partial\cdot K= 0$, the gauge-fixing parameter must be set to $c=\frac{2}{5}$ to remove logarithmic divergences (see Sec. II). By choosing $c=\frac{2}{5}$, however, $\psi$ does not correspond to a UIR of the dS group,
$$(Q_0 + 5)\psi=0,$$
and one encounters renowned difficulties in trying to quantize these fields with the so-called "imaginary mass" (with $c>-2$ or discrete series with $n>0$). The two-point functions for these fields demonstrate a pathological large distance behavior \cite{Ratra}
\begin{equation}\label{88}{\cal{W}}\approx|{\cal{Z}}(x,x')|^{-\frac{3}{2}+\frac{\sqrt{9+\frac{18}{2+c}}}{2}}.\end{equation}
The choice $c<-2$ removes this pathological behavior for the conformal sector, but a logarithmic divergence will appear in the traceless part.

In this paper, however, we proved that by applying the extra condition $\bar\partial\cdot K= 0$ (see Eq. (\ref{4.3}) and its underlying identities), one can eliminate the logarithmic divergence without fixing $c$ to the value of $2/5$. Therefore, respecting our calculations, in consistency with the dS physical representations, one can choose $c<-2$ to remove this pathological behavior from the pure-trace part, while the removal of the logarithmic divergence is assured. Once again "... our result is not in contradiction with the representations of the dS group. In our work, applying the aforementioned condition contracts the space of solutions without losing the minimal requirements for the solutions. While, the gauge fixing parameter must be set to $2/5$ when solution space is intact to remove the logarithmic singular terms."

Respecting the above capabilities, now, let us reconsider the situation. Comparing Eq. (\ref{66}) with (\ref{77}) reveals that in order to associate the scalar field $\psi$ to the scalar discrete series UIR of the dS group, we face the following cases:
\begin{itemize}
\item {For $n>0$ and $n<-3$ in Eq. (\ref{77}), the scalar field $\psi$ satisfying Eq. (\ref{66}) only corresponds to the dS UIR with $c>-2$. However, as already mentioned in this case, the associated two-point function presents a pathological large distance behavior \cite{Ratra}.}

\item{For $n=0,-3$ in Eq. (\ref{77}), obviously, there is no definite value for $c$ to relate the scalar field $\psi$ satisfying Eq. (\ref{66}) to the dS UIR.}

\item{For $n=-1,-2$ in Eq. (\ref{77}), interestingly, there exists one and only one "optimal" value for the gauge fixing parameter, for which, $\psi$ satisfying Eq. (\ref{66}) corresponds to the dS discrete series UIR, i.e. $c=-8$. Considering this choice, the associated two-point function is also free of any pathological large distance behavior (see (\ref{88})).}
\end{itemize}

In this regard, considering $c=-8$, Eq. (\ref{66}) converts to
\begin{equation}(Q_0 -2)\psi=0.\end{equation}
It, $\psi$, is indeed the conformally coupled massless scalar field in dS space, for which the corresponding two-point function is \cite{Chernikov109}
$$ {\cal W}^{(Hilbert)}_{cc}({\cal{Z}})= -{\frac{H^2}{8\pi^2}}\Big[{\frac{1}{1-\cal{Z}}}-i\pi\epsilon (x^0-x'^0)\delta(1-{\cal Z})\Big].$$
It is worth mentioning that the above two-point function has been calculated through the standard (Hilbert) quantization method. It preserves the dS invariance. However, as already discussed in detail, the covariant quantization of the gravitational part (traceless part) requires an indefinite metric field quantization based on the Krein quantization method. Therefore, here in order to preserve the self-consistency of the theory, the calculations are performed in the Krein context, as well. The two-point function, then, would be \cite{Gazeau1415}
$$ {\cal W}^{(Krein)}_{cc}({\cal{Z}})= {\frac{iH^2}{8\pi}}\epsilon (x^0-x'^0)\delta(1-{\cal Z}).$$

\section{Concluding remarks}
%\label{sec:}
In this paper, we have investigated the linear gravity (graviton-graviton interaction is absent) in de Sitter space. Pursuing the universal procedure, the result has been presented in terms of a spin-two (traceless) part and a gauge-dependent spin-zero (pure-trace) part.

\subsection{The spin-two part:\\ Covariant Gupta-Bleuler and Krein quantization}
Adapting the ambient space formalism and Gupta-Bleuler triplets, and in analogy with the standard calculations, we have shown that the spin-two sector of the theory can be written in terms of a projection tensor field and the minimally coupled massless scalar field. Thanks to a new version of the canonical quantization method by dropping the positivity requirement, the so-called Krein space quantization, we have been able to obtain the fully dS-covariant Wightman two-point function for this part. It is remarkably free of the pathological large distance behavior.

Here, we must underline that our result is not in contradiction with the standard calculations, for which, it is generally accepted that the phenomenon of de Sitter breaking is universal and the corresponding two-point function suffers from infrared divergences \cite{Mora53,Higuchi4317}. To see the points, it should be noted that the two-point function (\ref{6.25}) is written in terms of the massless minimally coupled two-point function ${\cal W}_{mc}(x,x')$ that is obtained through the Krein quantization scheme, in which, the one-particle sector is not a Hilbert space and the vacuum is dS invariant. While, the standard computations, through a Fock construction based upon a one-particle sector of the Hilbert space structure (equipped with a positive definite inner product), allow no de Sitter-invariant vacuum state for the minimally coupled field because of IR divergences \cite{Allen3136,Allen3771}. Indeed, through the standard computations, not only dS invariance but also gauge invariance is lost (since the Hilbert space of the states is noninvariant under gauge transformation $\phi\rightarrow\phi + \mbox{'constant'}$; See section V).\footnote{This problem is deeply analogous to the case of the massless scalar field in Minkowski space, for which a positive definite, analytic, Lorentz invariant, and local two-point function is not accessible. More exactly, if one insists on Lorentz-invariance necessarily lose positivity.} So it is not surprising that the calculated graviton two-point function (the spin two sector) based on it breaks the dS invariance. Of course, even in our calculations if one abandons the Krein quantization scheme for the minimally coupled field and use the normal modes to construct the two-point function, in agreement with standard computations \cite{Mora53,Higuchi4317}, the dS invariance is lost.

In spite of the presence of negative norm modes in the Krein context, no negative energy can be measured; the energy operator has positive expectation values in all physical states which assures a reasonable physical interpretation of the theory (the method indeed fulfils the so-called Wald axioms) \cite{de Bievre6230,Gazeau1415}.

\subsection{The spin-zero part:\\ Consistent gauge-fixing procedure}
We have also perused the pure-trace part of the theory. It has been proved that if considerations of the rigorous group theoretical approach to the subject are taken into account, through the suitable gauge-fixing procedure, the pure-trace part is written in terms of the conformally coupled massless scalar field. Indeed, de Sitter invariance is well preserved, and the theory is free of any IR divergences. Therefore, the obtained result for the spin-zero part is in complete agreement with the usual viewpoint which asserts that this part is gauge dependent and, hence, the presented divergences can be suppressed by a proper gauge-fixing scheme \cite{Mora53}. In addition, this result is remarkably consistent with a requirement of the strong equivalence principle; in all metric theories of gravity, including general relativity, in which the scalar field is not part of the gravitational sector (like our studied case), the coupling constant should be conformal in order for the short distance propagators of the theory to be compatible with those found in a Minkowski spacetime \cite{Faraoni53,Sonego10}.
$$$$

\subsection{Outlook:\\ Interacting Krein QFT in curved space}
In order to include the interacting cases in the presence of the quantum effects of gravity, the theory requires more investigations. As a matter of fact, through the Krein method, the symmetric two-point function (Hadamard function) vanishes, and so, it cannot have the Hadamard property. Therefore, it seems that there would be a difficulty to construct perturbative interacting QFT in curved spacetime \cite{Brunetti,Hollands}. In this regard, we should emphasize that although in this quantization scheme the vacuum is unique and does not determine the physical states space, the link between the physical space and the two-point function (a function with Hadamard property but with another meaning) remains \cite{Gazeau1415,Garidi}. Nevertheless, the construction of an interacting Krein QFT in curved space remains open.

\section*{Acknowledgements}
We would like to thank Professor J. P. Gazeau for remarkable comments and providing us with his unpublished notes \cite{Gazeau0}. We would also like to thank the referee for his/her useful comments.

\begin{appendix}
\setcounter{equation}{0}
\section{Some useful relations}
Some useful identities are collected in this appendix:
\begin{equation} \partial_2 \cdot \theta \phi = -H^2D_1\phi, \end{equation}
\begin{equation} Q_2 D_2 K=D_2Q_1K,\;\;Q_2\theta\phi=\theta Q_0\phi,\end{equation}
\begin{equation} (Q_0-2)x=x Q_0 -6x -2H^{-2}\bar\partial,\end{equation}
\begin{equation} \bar\partial(Q_0-2) = Q_0\bar\partial -8\bar\partial-2H^2Q_0 x - 8H^2x,\end{equation}
\begin{equation} [Q_0Q_2,Q_2Q_0]{{\cal{K}}}= 4S(x - \bar\partial)\bar\partial{\cdot{\cal{K}}},\end{equation}
\begin{equation} Q_2S\bar Z K = S\bar Z (Q_1-4)K - 2H^2D_2x\cdot Z K + 4\theta Z\cdot K.\end{equation}

The following relations are considered to calculate the two-point function:
\begin{eqnarray}
\bar{\partial}_\alpha f({\cal{Z}})=-(x'\cdot\theta_{\alpha})\frac{d f(\cal{Z})}{d{\cal{Z}}},
\end{eqnarray}
\begin{eqnarray}
\theta^{\alpha\beta}\theta'_{\alpha\beta}=\theta\cdot\cdot\theta'=3+{\cal{Z}}^2,
\end{eqnarray}
\begin{eqnarray}
(x.\theta'_{\alpha'})(x\cdot\theta'^{\alpha'})={\cal{Z}}^2-1,
\end{eqnarray}
\begin{eqnarray}
(x.\theta'_{\alpha})(x'\cdot\theta^{\alpha})={\cal{Z}}(1-{\cal{Z}}^2),
\end{eqnarray}
\begin{eqnarray}
\bar{\partial}_\alpha(x\cdot\theta'_{\beta'})=\theta_{\alpha}\cdot\theta'_{\beta'},
\end{eqnarray}
\begin{eqnarray}
 \bar{\partial}_\alpha(x'\cdot\theta_{\beta})=x_\beta(x'\cdot\theta_{\alpha})-{\cal{Z}}\theta_{\alpha\beta},
\end{eqnarray}
\begin{eqnarray}
\bar{\partial}_\alpha(\theta_{\beta}\cdot\theta'_{\beta'})=x_\beta(\theta_{\alpha}\cdot\theta'_{\beta'})+ \theta_{\alpha\beta}(x\cdot\theta'_{\beta'}),
\end{eqnarray}
\begin{eqnarray}
\theta'^{\beta}_{\alpha'}(x'\cdot\theta_{\beta})=-{\cal{Z}}(x\cdot\theta'_{\alpha'}),
\end{eqnarray}
\begin{eqnarray}
\theta'^{\gamma}_{\alpha'}(\theta_{\gamma}\cdot\theta'_{\beta'})=\theta'_{\alpha'\beta'}+(x\cdot\theta'_{\alpha'})(x\cdot\theta'_{\beta'}),
\end{eqnarray}
\begin{eqnarray}\label{A.16}
Q_0f({\cal{Z}})=(1-{\cal{Z}}^2)\frac{d^2 f(\cal{Z})}{d{\cal{Z}}^2}-4{\cal{Z}}\frac{d f(\cal{Z})}{d{\cal{Z}}}.
\end{eqnarray}

\setcounter{equation}{0}
\section{Mathematical relations underlying the Eq. (\ref{4.14})}
Considering the identities given in the Appendix A, the conditions $x\cdot K=\bar\partial\cdot K= 0$ and $Q_0K=0$, we have \cite{Dehghani064028}
\begin{equation} \label{4.12} (Q_1+6) D_1 (Z_1\cdot K) = 6 D_1 (Z_1\cdot K),\end{equation}
\begin{equation} \label{4.13} (Q_1+6) x (Z_1\cdot K) = 6 x (Z_1\cdot K),\end{equation}
\begin{equation} \label{4.11} (Q_1+6)Z_1\cdot\bar\partial K = 6Z_1\cdot\bar\partial K+ 2H^2 D_1 (Z_1\cdot K),\end{equation}
\begin{equation} \label{4.10'} (Q_1+6)[H^2(x\cdot Z_1)K] = 2\Big[ H^2x(Z_1\cdot K) - Z_1\cdot\bar\partial K \Big].\end{equation}

Combining (\ref{4.13}), (\ref{4.11}), and (\ref{4.10'}) leads to
\begin{equation} H^2(x\cdot Z_1)K =\frac{1}{3} \Big[ \frac{1}{3} H^2 D_1(Z_1\cdot K) + H^2 x (Z_1 \cdot K) - Z_1\cdot\bar\partial K \Big].\end{equation}

With regard to Eqs. (\ref{4.12}) and (\ref{4.10'}), we then obtain
\begin{equation} \label{4.10} (Q_1+6)\Big[ \frac{1}{9}  D_1(Z_1\cdot K) + (x\cdot Z_1)K \Big]= 6(x\cdot Z_1)K.\end{equation}
Respecting these identities, one can easily obtain Eq. (\ref{4.14}).

\setcounter{equation}{0}
\section{Mathematical relations underlying the Eq. (\ref{A.7})}
Substituting (\ref{A.1}) into Eq. (\ref{4.8}) leads to
\begin{equation}\label{A.2} K_\alpha = -\frac{\sigma}{2}\Big[\bar Z_{2\alpha} + (\sigma + 2)\frac{(x\cdot Z_2)}{(x\cdot\xi)}\bar\xi_\alpha\Big]\phi. \end{equation}
Using this equation, we have
\begin{equation}\label{A.3}
Z_1\cdot K =-\frac{\sigma}{2}\Big[ H^2(\sigma+3)(x\cdot Z_1)(x\cdot Z_2) + (Z_1\cdot Z_2)\Big]\phi,
\end{equation}
\begin{eqnarray}\label{A.4}
Z_1\cdot\bar\partial K_\beta  = - \frac{\sigma}{2} H^2 \Big(\Big[Z_1\cdot Z_2 \hspace{3.5cm}\nn\\
+ H^2(\sigma+3)(x\cdot Z_1)(x\cdot Z_2)\Big]x_\beta \hspace{1cm}\nn\\
+ (\sigma+3)(x\cdot Z_2)\bar Z_{1\beta} + \sigma(x\cdot Z_1)\bar Z_{2\beta}\hspace{1cm} \nn\\ +
(\sigma + 2)\Big[H^{-2}\frac{Z_1\cdot Z_2}{x\cdot\xi} + \sigma\frac{(x\cdot Z_1)(x\cdot Z_2)}{x\cdot\xi} \Big]\bar\xi_\beta\Big)\phi,\hspace{0.5cm}
\end{eqnarray}
\begin{eqnarray}\label{A.5}
D_{1\beta}(Z_1\cdot K) = -\frac{\sigma}{2} \Big((\sigma+3)\Big[ (x\cdot Z_2)\bar Z_{1\beta} + (x\cdot Z_1)\bar Z_{2\beta}\Big]\nn\\
+ \sigma\Big[H^{-2}\frac{Z_1\cdot Z_2}{x\cdot\xi} + (\sigma+3)\frac{(x\cdot Z_1)(x\cdot Z_2)}{x\cdot\xi} \Big]\bar\xi_\beta\Big)\phi.\hspace{0.5cm}
\end{eqnarray}
Note that, for simplicity, the conditions $Z_1\cdot\xi= Z_2\cdot\xi=0$, are imposed. Because of these conditions, the degree of freedom of 5-vectors $Z_1$ and $Z_2$ then is reduced from 5 to 4.

Substituting (\ref{A.2})-(\ref{A.5}) into (\ref{4.14}) leads to
$$ K_{g\beta} = \frac{\sigma cH^2}{12(c-1)}\Big[{g_1(c,\sigma)}(x\cdot Z_2)\bar Z_{1\beta}+ {g_2(c,\sigma)}(x\cdot Z_1)\bar Z_{2\beta}$$
$$ + \Big({g_3(c,\sigma)} H^{-2}\frac{Z_1\cdot Z_2}{x\cdot\xi}+ {g_4(c,\sigma)}\frac{(x\cdot Z_1)(x\cdot Z_2)}{x\cdot\xi} \Big)\bar\xi_\beta \Big]\phi,$$
where
$$ {g_1(c,\sigma)} = 2(\sigma+3)\frac{1-4c}{9c},$$
$${g_2(c,\sigma)} = (\sigma+3)\frac{2+c}{9c} + \frac{2-5c}{c} - \sigma,$$
$$ {g_3(c,\sigma)} = \sigma \frac{2+c}{9c} - (\sigma+2), $$
$${g_4(c,\sigma)} = \sigma (\sigma+3)\frac{2+c}{9c} + (\sigma+2)\frac{2-5c}{c} - \sigma(\sigma+2).$$
Then, we have
\begin{eqnarray}\label{A.6}
D_{2\alpha}K_{g\beta} = \frac{c\sigma}{12(c-1)} {\cal S} \Big[ {g'_1(c,\sigma)} \bar Z_{1\alpha}\bar Z_{2\beta}\hspace{2.5cm}\nn\\
+ {g'_2(c,\sigma)} \frac{(x\cdot Z_2)}{(x\cdot\xi)}\bar{Z}_{1\alpha}\bar\xi_\beta + {g'_3(c,\sigma)} \frac{(x\cdot Z_1)}{(x\cdot\xi)}\bar{Z}_{2\alpha}\bar\xi_\beta \hspace{0.7cm}\nn\\
+ \Big( {g'_4(c,\sigma)} Z_1\cdot Z_2 + {g'_5(c,\sigma)} H^2 (x\cdot Z_1)(x\cdot Z_2) \Big) \theta_{\alpha\beta}\hspace{0.5cm} \nn\\
+ (\sigma-1)\Big( {g'_4(c,\sigma)} H^{-2} \frac{Z_1\cdot Z_2}{(x\cdot\xi)^2}\hspace{3.9cm}\nn\\
+ {g'_6(c,\sigma)} \frac{(x\cdot Z_1)(x\cdot Z_2)}{(x\cdot\xi)^2} \Big) \bar\xi_\alpha\bar\xi_\beta \Big]\phi,\hspace{0.5cm}
\end{eqnarray}
where
$$ {g'_1(c,\sigma)} = 2(\sigma+3)\frac{2+c}{9c}+  \frac{2-5c}{c} - (2\sigma+3),$$
$$ {g'_2(c,\sigma)} = 2\sigma(\sigma+3)\frac{2+c}{9c}+ (\sigma+2)\frac{2-5c}{c} - \sigma(2\sigma+5),$$
$$ {g'_3(c,\sigma)} = 2\sigma(\sigma+3)\frac{2+c}{9c} + 2(\sigma+2)\frac{2-5c}{c} - 2\sigma(\sigma+1),$$
$$ {g'_4(c,\sigma)} = \sigma\frac{2+c}{9c} - (\sigma+2),$$
$$ {g'_5(c,\sigma)} = (\sigma+3)\Big[(\sigma+2)\frac{2+c}{9c} + \frac{2-5c}{c} - (\sigma+1)\Big],$$
$$ {g'_6(c,\sigma)}=\sigma(\sigma+3)\frac{2+c}{9c}+(\sigma+2)\frac{2-5c}{c}-\sigma(\sigma+2).$$
Finally, considering the above identities and Eqs. (\ref{4.4}) and (\ref{4.1}), one can easily obtain (\ref{A.7}).

\setcounter{equation}{0}
\section{An interacting theory in Minkowski spacetime}
In this appendix, we briefly study the Krein space quantization behavior in Minkowski spacetime for a theory with interaction. In this regard, let us illustrate the points by giving a simple example: an interacting scalar field with the following Lagrangian density,
\begin{equation}{\cal{L}}=g^{\mu\nu}\partial_\mu\phi\partial_\nu\phi - m^2\phi - V(\phi).\end{equation}
Here, ${\cal{H}}={\cal{H}}_{+}\oplus {\cal{H}}_{-}$ determines the free-field Fock space, while the space of physical states ${\cal{H}}_{+}$ is closed and positive.

Regarding the appearance of unphysical states in the method, unitarity of the theory would be preserved by the following procedure, which is the so-called unitarity condition: let $\Pi_+$ be the projection over ${\cal{H}}_+$,
\begin{equation}\Pi_+ = \sum_{\{\alpha_+\}} |\alpha_+><\alpha_+|,\;\;\;\;\; |\alpha_+>\;\in {\cal{H}}_+,\end{equation}
so that
\begin{equation}\label{unicon} \Pi_+ \phi \Pi_+ |\alpha > = \left\{\begin{array}{rl} &\phi_+ |\alpha >, \;\;\;\;\;\; \mbox{if}\; |\alpha>\;\in {\cal{H}}_+ \vspace{2mm}\\\vspace{2mm} &0,\;\;\;\;\;\;\;\;\;\;\;\;\;\;\;\;\mbox{if}\; |\alpha>\;\in {\cal{H}}_{-} \\\end{array}\right. \end{equation}
On this basis, then, the same predictions as the usual scalar field theory would be assured by substituting the standard choice for the Lagrangian potential, i.e. $V(\phi)$, with $V'(\phi)\equiv V(\Pi_+\phi\Pi_+)$, which is the restriction of $V$ to the positive energy modes,
\begin{equation}{\cal{L}}=g^{\mu\nu}\partial_\mu\phi\partial_\nu\phi - m^2\phi - V'(\phi).\end{equation}

As accurately discussed in Ref. \cite{Garidi}, pursuing the same procedure for the other canonically quantizable theories, i.e. replacing the various fields $\chi$ (for interacting terms) by their restricted forms $\Pi_+^\chi \chi \Pi_+^\chi$, where $\Pi_+^\chi$ is the corresponding projector, the unitarity of the theory would be guaranteed. It should be noted that the so-called radiative corrections are the same as in the usual QFT. The only difference between such a Krein field theory and the usual ones is the vanishing of the vacuum energy of the free field. [To obtain a detailed construction of the quantization method and, in particular, the unitarity condition and compatibility with the (Hilbert space) QFT's counterpart in the Minkowskian limit, one could refer to Ref. \cite{Garidi}).]

The crucial point, which should be underlined here, is that the operator $\Pi_+$ may not exist in a curved spacetime. Therefore, on such a space, the difficulty of the interacting field, as is well known, is much more elaborate.
\end{appendix}

\end{document}